\documentclass[a4paper,10pt,twocolumn]{article} 

\usepackage{amsmath, amssymb, amsthm}
\usepackage[mathscr]{euscript}     
\usepackage{dsfont} 
\usepackage{graphicx} 
\usepackage{qip} 
\usepackage[continuous]{thmenvironments} 
\usepackage{layoutcommands} 

\usepackage{underscore} 
\usepackage[shortcuts]{extdash} 
\usepackage{authblk} 
\usepackage{appendix} 
\usepackage{microtype} 
\usepackage{enumitem} 
\setlist{nolistsep}
\usepackage[margin=10pt,font=small,labelfont=bf,labelsep=endash,hypcap=true]{caption}
\usepackage{tikz} 
\usetikzlibrary{arrows,fit,positioning,shapes}

\usepackage{hyperref} 
\usepackage[open,numbered]{bookmark} 
\usepackage{cite} 
\usepackage{hyperlinks} 
\usepackage{eprint} 

\hypersetup{
  pdftitle={Key recycling in authentication},
  pdfauthor={Christopher Portmann},
  pdfsubject={Cryptography},
  pdfstartview={FitH}
}

\newcommand{\auth}{\text{auth}}

\tikzset{sArrow/.style={->,>=stealth,thick}}
\tikzset{arrowLabel/.style={auto}}

\title{Key recycling in authentication}

\author{Christopher Portmann\email{chportma@phys.ethz.ch}}

\affil{Institute for Theoretical Physics, ETH Zurich, 8093 Zurich, Switzerland.}

\date{}

\begin{document}

\maketitle

\begin{abstract}
In their seminal work on authentication, Wegman and
Carter propose that to authenticate multiple messages, it
is sufficient to reuse the same hash function as long as each tag is
encrypted with a one-time pad. They argue that because the one-time
pad is perfectly hiding, the hash function used remains completely
unknown to the adversary.

Since their proof is not composable, we revisit it using a composable
security framework. It turns out that the above argument is
insufficient: if the adversary learns whether a corrupted message was
accepted or rejected, information about the hash function is leaked,
and after a bounded finite amount of rounds it is completely known. We
show however that this leak is very small: Wegman and Carter's
protocol is still $\eps$\=/secure, if $\eps$\=/almost strongly
universal$_2$ hash functions are used. This implies that the secret
key corresponding to the choice of hash function can be reused in the
next round of authentication without any additional error than this
$\eps$.

We also show that if the players have a mild form of synchronization,
namely that the receiver knows when a message should be received, the
key can be recycled for any arbitrary task, not only new rounds of
authentication.
\end{abstract}


\section{Introduction}
\label{sec:intro}

If a player, say, Bob, receives a message $x$ that claims to come from
Alice, he might wish to know if this is true, or if the message was
generated or modified by some adversary. This task is called
\emph{authentication}, and in their seminal work~\cite{WC81}, Wegman
and Carter showed that it can be achieved with information\-/theoretic
security by appending a tag $t$ to the message (often called a message
authentication code or MAC), where $t = h_k(x)$, $\cH = \{h_k\}_{k \in
  \cK}$ is a family of almost strongly universal$_2$ (ASU$_2$) hash
functions,\footnote{ASU$_2$ hashing was only formally defined later by
  Stinson~\cite{Sti94}. A family of functions is said to be ASU$_2$ if
  any two different messages are almost uniformly mapped to all pairs
  of tags. An exact definition is given in \defref{def:universalhash}
  on \pref{def:universalhash}.} and $k$ is a secret key shared by
Alice and Bob.

Wegman and Carter~\cite{WC81} propose a scheme to use fewer bits of
key when multiple messages are to be authenticated: each tag should be
encrypted with a fresh one-time pad (OTP), but the same hash function
can be used each time. Alice thus appends the tag $t_i = h_{k_1}(x_i)
\xor k_2^i$ to her \ith{i} message $x_i$, where $k_1$ is used for all
messages and $k^i_2$ is a fresh key used only in this
round. 

To prove the security of this scheme, Wegman and Carter show that
given any number of message\-/tag pairs $(x_1,t_1),(x_2,t_2),\dotsc,$
the secret key $k_1$ is still perfectly uniform. They then argue that
the probability of an adversary successfully corrupting any new
message is the same in every round, and guaranteed to be small by the
properties of the ASU$_2$ hash functions.

However, proving that a protocol is secure in a stand\-/alone model
does not necessarily guarantee that it is still secure when combined
with other protocols, not even when combined with itself like Wegman
and Carter's scheme. A lot of research has gone into composability of
cryptographic tasks in recent years. A general framework for proving
composable security was developed by
Canetti~\cite{Can01,Can13,CDPW07}, and dubbed \emph{Universally
  Composable (UC) security}. Independently, Backes, Pfitzmann and
Waidner~\cite{PW00,PW01,BPW04,BPW07} introduced the equivalent
notion of \emph{Reactive Simulatability}. These security notions have
been extended to the quantum setting by Ben-Or and Mayers~\cite{BM04}
and Unruh~\cite{Unr04,Unr10}. More recently, Maurer and Renner
proposed a new composable security framework~\cite{MR11},
\emph{Abstract Cryptography (AC)}, which both generalizes and
simplifies previous frameworks. It defines composability of abstract
systems, without specifying the underlying computational model, and is
thus equally valid for classical and quantum security.\footnote{This
  unified treatment of classical and quantum cryptography allows for a
  seamless composition of the two, see \remref{rem:ac.lifting} on
  \pref{rem:ac.lifting}.} We use AC in this work, because of the
extra clarity it provides. The same results could however be obtained
from other composable security frameworks.

An essential application of information\-/theoretic authentication is
in quantum key distribution (QKD) protocols.\footnote{We refer to a
  review article such as \cite{SBCDLP09} for a general overview of
  QKD.} Every (classical) message exchanged between the two parties
generating the key needs to be authenticated with
information\-/theoretic security in order to guarantee that the
composed protocol remains secure in the presence of a computationally
unbounded adversary. Recycling the hash function is a practical way to
save a large part of the secret key consumed in each round. Ben-Or et
al.~\cite{BHLMO05} and M\"uller-Quade and Renner~\cite{MR09} discuss
the composability of QKD. More recently, Portmann and
Renner~\cite{PR14} provide an extensive review of composable security
with many examples from QKD. In particular, they show that if the
authentication and QKD schemes are proven to be composable, and if a
short initial key is available, a continuous key stream can be
generated with arbitrarily small error. But since Wegman and Carter's
security proof does not fit in any composable security framework, it
raises the question of whether this QKD application is still secure.

\subsection{Related work}

Many works, e.g.,~\cite{Kra94,Kra95,Rog99,AS96}, reuse Wegman and
Carter's authentication scheme with key recycling, and they all sketch
the security using the same non\-/composable argument as Wegman and
Carter. Composable security for key recycling in authentication has
been studied in the case of quantum messages by Hayden, Leung and
Mayers~\cite{HLM11},\footnote{In the quantum case, recycling is
  possible due to the no-cloning theorem. If the message is
  successfully authenticated, the eavesdropper cannot have any
  information about it or the corresponding cipher, and therefore has
  no information about the secret key used either.} but to the best of
our knowledge has not been treated when the messages are
classical.\footnote{Standard information\-/theoretic authentication
  \--- in which a different hash function is used for every new
  message \--- has not been explicitly proven to be composable
  either. But as we note in the Appendix, its security reduces to the
  stand\-/alone security criterion used in the literature, and is
  therefore immediate from Stinson's work~\cite{Sti94}.}
Computationally secure variants of Wegman and Carter's scheme have
been proposed~\cite{Sho96,Ber05}, but not analysed in a composable
framework either.\footnote{Though the composability of public\-/key
  authentication constructed from digital signatures has been
  extensively studied~\cite{Can04}.}

Some works \cite{CL08,AL11} have pointed out that
information\-/theoretic authentication might not be composable with
QKD. They attempt to study this problem by analyzing the security of
authentication when the secret keys used are not perfect. In
particular, Abidin and Larsson~\cite{AL11} suggest that when QKD and
authentication with key recycling are combined recursively, and the
(imperfect) secret key resulting from QKD is fed back into the next
round of authentication and QKD, the total error could increase
exponentially in the number of rounds.

\subsection{New results}

We therefore prove that Wegman and Carter's authentication scheme with
key recycling~\cite{WC81} is secure using the AC framework of Maurer
and Renner \cite{MR11}. Since this framework defines composable
security independently from the computational model of the underlying
system, any (classical) protocol proven to be secure is immediately
composable with quantum protocols. The authentication scheme with key
recycling studied in this work can thus be used by a QKD protocol.

We show that simply by learning if a message was accepted, the
recycled hash function is gradually leaked to the adversary, even when
the key used for the OTP is perfect. This leakage is however very
small: we prove that this scheme is indeed \emph{$\eps$\=/secure} if
the hash functions used are $\eps$\=/ASU$_2$. Since for any good
ASU$_2$ hash function construction $\eps$ decreases exponentially fast
in the size of the family, $\log |\cH|$, it can easily be made
arbitrarily small. As a consequence, the doubts of \cite{CL08,AL11}
are unfounded for all ranges of the parameter $\eps$.  In fact, the
hash functions used are slightly weaker than ASU$_2$ hashing, namely
almost XOR universal$_2$ (AXU$_2$) hash functions.\footnote{A family
  of functions is said to be AXU$_2$ if the bitwise XOR of the hash of
  any two different messages is almost uniform over the choice of hash
  function. See \defref{def:xorhash} on \pref{def:xorhash} for an
  exact definition.}

In a composable framework, a protocol is said to be
\emph{$\eps$\=/secure}, if every use of this protocol increases the
overall error by at most $\eps$.\footnote{More precisely, a protocol
  is $\eps$\=/secure if the resulting real system is $\eps$\=/close to
  an ideal system (see \defref{def:ac.security} on
  \pref{def:ac.security}). By a triangle inequality type argument, the
  distance of a composed protocol from ideal is the sum of the
  distances of the individual protocols~\cite{PR14}.} An immediate
implication of our result and the composability of the AC framework is
that if $n$ messages are authenticated this way in each round of an
$\eps'$\=/secure QKD protocol, which is run $r$ times, recycling the
same hash function throughout all the runs, the concatenation of all
generated keys \--- all bits of key that have not be used by other
rounds of QKD \--- has distance at most $rn\eps+r\eps'$ from uniform.

We analyze Wegman and Carter's authentication scheme in two
settings. In the first, the sender and receiver, Alice and Bob, have a
mild form of synchronization: Bob knows that a message has been
sent,\footnote{A similar notion of synchronization was used by
  Maurer~\cite{Mau13} to authenticate a long message given an
  authentic channel for a short message, but no shared key.} and
rejects any message that is delayed for too long, even if it passes
the authentication. The reason for this is that, were Alice to reuse
the key in an application which leaks it to Eve \--- e.g., encrypting
a message known to Eve \--- Eve could then use this information about
the key to modify the original message and only then deliver it to
Bob, breaking the authentication scheme. So Alice only recycles her
key \emph{after} the timeout has occurred.\footnote{Alternatively,
  Alice could recycle her key before the timeout if she receives a
  confirmation of reception from Bob. But this requires extra
  authenticated communication, see the discussion in
  \secref{sec:sync}.}

The second setting is a completely asynchronous network. Here, the key
cannot be reused arbitrarily. However, we show that it is still safe
to recycle it for further rounds of authentication. This is because
obtaining extra message\-/tag pairs give no information to Eve about
the recycled key. So the delayed attack outlined above is useless.

\subsection{Key consumption}

Finding an authentication scheme which allows part of the key to be
recycled is quite trivial: if only part of the key is actually used by
the protocol, the remaining key bits can be ``reused''. Alternatively,
the classical message could be authenticated using the quantum scheme
of Hayden et al.~\cite{HLM11}.\footnote{Although this would only allow
  the key to be recycled if the authentication is successful.} The
former obviously does not allow any key to be saved, the latter is
impossible to implement (with today's technology).

Wegman and Carter's key recycling scheme is of interest, because it
results in a net gain in secret key bits consumed and is efficiently
implementable (if the chosen hash functions are efficiently
implementable). It is already in use in QKD systems, for example in
\cite{Wal14}. There, the authors chose to use an ASU$_2$ hash function
construction of Bierbrauer et al.~\cite{BJKS94}, which is efficiently
implementable in hardware. This ASU$_2$ family \--- which has size
$\log |\cH| \approx 2 \log \log |\cX| + 3 \log |\cT|$, where $\cX$ is
the message alphabet and $\cT$ the tag alphabet \--- is equivalent to
an AXU$_2$ family with an additional OTP, i.e., it already has the
form $h_{k_1,k_2}(x) = h'_{k_1}(x) \xor k_2$. It can thus be used in a
scheme with key recycling without needing any extra key bits to
encrypt the tag for the first message. All subsequent messages then
only need $\log |\cT|$ bits of fresh key, at least a $2/3$ gain over
using a new hash function each time.

\subsection{Structure of this paper}

We start in \secref{sec:tech} with some brief technical
preliminaries. In \secref{sec:ac} we introduce Abstract Cryptography,
and model standard authentication (without key recycling) in the AC
framework to illustrate its basic principles. In \secref{sec:sync} we
define the task of recycling part of the key for arbitrary use. We
show that this is achieved by Wegman and Carter's scheme with a mild
form of synchronization between sender and receiver. In
\secref{sec:async} we recursively use the authentication scheme $n$
times, but in a fully asynchronous network, and prove that it is
secure with error $n\eps$. And finally in \secref{sec:leak} we take a
closer look at the secret key which is leaked to the adversary, and
show that an optimal attack over $n$ rounds of authentication which
takes advantage of this key leakage allows the adversary to break the
scheme with probability exactly $n\eps$.


\section{Technical preliminaries}
\label{sec:tech}

\subsection{Notation}
\label{sec:notation}

In the AC framework that we introduce in the next section, the real
and ideal systems\footnote{Composable security frameworks define
  security as the distance between the real system and some ideal
  system that performs the task in an perfect way.} are modeled as
(black) boxes with ports that take inputs and produce outputs. We use
upper case letters, \--- $X$, $Y$ \--- to label these ports, lower
case letters \--- $x$, $y$ \--- to denote the values that are in- or
output, and a calligraphic font for the alphabets of permitted values
\--- $\cX$, $\cY$. To describe probability distributions over these
values we write $P_X$, $Q_X$, where the subscript denotes the
port. Different distributions over the same port are differentiated by
the main upper case letter, e.g., $P$, $Q$. For example, in the real
setting $x$ might be output at port $X$ with probability $P_X(x)$ and
in the ideal setting with probability $Q_X(x)$. To denote the
probability of $y$ being output at port $Y$ given that $x$ was input
at port $X$, we use standard notation and write $P_{Y|X}(y|x)$.

\subsection{Statistical distance}
\label{sec:sd}

To measure the distance between two settings with distributions $P_X$
and $Q_X$, we use the \emph{statistical} or \emph{total variation}
distance, which is given by
\begin{align}
    & \frac{1}{2} \sum_{x \in \cX} \left| P_X(x) - Q_{X}(x) \right|
    \notag \\
    & \qquad \qquad =
    \max_{\cX' \subseteq \cX} \sum_{x \in \cX'} \left(
      P_X(x) - Q_{X}(x) \right) \notag \\
    & \qquad \qquad = \sum_{x : P_X(x) > Q_{X}(x)} \left( P_X(x) -
    Q_{X}(x) \right). \label{eq:sd} 
\end{align}

In this paper we use many times this last equality to bound the
statistical distance. We find the subset $\cX' \subseteq \cX$ for
which $P_X(x) \geq Q_{X}(x)$, and then only need to evaluate the
distributions $P_X$ and $Q_X$ on this subset.

\subsection{Universal hashing}
\label{sec:universalhash}

Standard authentication is performed by sending the message along with
a hash of it, where the hash function is taken from an ASU$_2$ family,
defined as follows.

\begin{deff}[strongly universal$_2$ hash
  function\footnoteremember{fn:universalhash}{The more common
    definition of strongly universal$_2$
    hashing~\cite{Sti94,BJKS94,Sti95,AS96} has an extra condition,
    namely that for all $x \in \cX$ and $t \in \cT$, $\Pr \left[
      h_k(x) = t \right] = \frac{1}{|\cT|}$. This is however not a
    necessary condition to prove the security of authentication, so we
    omit it.}~\cite{Sti94}]
  \label{def:universalhash}
  A family of hash functions $\{h_k : \cX \to \cT \}_{k \in \cK}$ is
  said to be $\eps$-almost strongly universal$_2$ ($\eps$-ASU$_2$) if for
  $k$ chosen uniformly at random and all $x_1,x_2 \in \cX$ with $x_1
  \neq x_2$ and all $t_1,t_2 \in
  \cT$, \begin{equation} \label{eq:universalhash} \Pr_k \left[ h_k(x_1)
      = t_1 \text{ and } h_k(x_2) = t_2 \right] \leq
    \frac{\eps}{|\cT|}. \end{equation}
\end{deff}

In the case of key recycling, we use a weaker family of hash
functions, but then encrypt the tag with a OTP. These hash functions
have been dubbed \emph{$\eps$\=/almost XOR universal$_2$} by
Rogaway~\cite{Rog99}, \emph{$\eps$\-/otp secure} by
Krawczyk~\cite{Kra94,Kra95}, and \emph{$\eps\text{\=/}\Delta$ universal}
by Stinson~\cite{Sti95}.\footnote{Stinson~\cite{Sti95} generalizes this
  notion to any additive abelian group $\cT$ instead of only bit
  strings.}

\begin{deff}[XOR universal$_2$ hash function~\cite{Rog99}]
  \label{def:xorhash}
  A family of hash functions $\{h_k : \cX \to \cT \}_{k \in \cK}$ for
  $\cT = \{0,1\}^m$ is said to be $\eps$-almost XOR universal$_2$
  ($\eps$-AXU$_2$) if for $k$ chosen uniformly at random and all
  $x_1,x_2 \in \cX$ with $x_1 \neq x_2$ and all $t \in
  \cT$, \begin{equation} \label{eq:xorhash} \Pr_k \left[ h_k(x_1) \xor
      h_k(x_2) = t \right] \leq \eps. \end{equation}
\end{deff}

It is immediate from this definition that the hash function
$g_{k_1,k_2}(x) \coloneqq h_{k_1}(x) \xor k_2$ is $\eps$-ASU$_2$,
i.e., for all $x_1,x_2 \in \cX$ with $x_1 \neq x_2$ and all $t_1,t_2
\in \cT$, \[\Pr_{k_1,k_2} \left[ g_{k_1,k_2}(x_1) = t_1 \text{ and }
  g_{k_1,k_2}(x_2) = t_2 \right] \leq \frac{\eps}{|\cT|}.\]

We derive two more useful statements about hash functions of the form
$g_{k_1,k_2}(x) = h_{k_1}(x) \xor k_2$, where $h_{k_1}$ is
$\eps$-AXU$_2$. The first is that the tag for one message is uniformly
distributed and independent from the key $k_1$: since XORing a uniform
string $k_2$ to any value yields a uniform string we have for all $x$,
$t$ and $k_1$, \begin{equation} \label{eq:xorhash1} \Pr_{k_2} \left[
    g_{k_1,k_2}(x) = t \right] = \frac{1}{|\cT|}.\end{equation} The
second is a conditional form of the strongly universal$_2$ property of
these hash functions: by combining the two equations above we
immediately get \begin{equation} \label{eq:xorhashotp} \Pr_{k_1,k_2}
  \left[ g_{k_1,k_2}(x_2) = t_2 \middle| g_{k_1,k_2}(x_1) = t_1
  \right] \leq \eps. \end{equation}

These probabilities of hashing to a certain value can be rewritten has
as conditional probabilities of the in- and outputs of an
authentication system in the following way. If a message $x$ is input
into an authentication system which outputs a tag $t = h_k(x)$, the
probability distribution of the tag is given by \[P_{T|X}(t|x) =
\Pr_{k} \left[ h_{k}(x) = t \right].\] If an adversary obtains a valid
message and tag $x\|t$ and chooses some $x'\|t'$ to input into a
verification system which outputs a decision $y \in \{\text{\tt
  acc},\text{\tt rej}\}$, the probability of this corrupted message of
being accepted is
\begin{multline*}P_{Y|XTX'T'}(\text{\tt acc}|x,t,x',t') \\ = \Pr_{k} \left[ h_{k}(x')
  = t' | h_{k}(x) = t \right].\end{multline*}



\section{Abstract cryptography}
\label{sec:ac}

To model security we use Maurer and Renner's~\cite{MR11} Abstract
Cryptography framework. In this section we give an introduction to the
special case needed in this work, namely information\-/theoretic
security of classical systems with three parties, an honest Alice and
Bob and dishonest Eve. For more extensive introductions to the AC
framework in the three party setting we refer to \cite{Mau12} and
\cite{PR14}. The first reference treats only the classical case, the
second is also valid for quantum systems.

The AC security definition\footnote{\defref{def:ac.security} on
  \pref{def:ac.security}.} applies to abstract systems, which can be
instantiated with different models of computation. In particular, it
is equally valid for classical and quantum systems~\cite{PR14,DFPR14},
and any protocol proven to be information\-/theoretic ``classically
secure'' is immediately ``quantum secure''. An equivalent to Unruh's
lifting lemma~\cite{Unr10} \--- which proves that classical UC
security of a classical scheme implies quantum UC security \--- is
unnecessary, since this is immediate from the model.\footnote{See
  \remref{rem:ac.lifting} on \pref{rem:ac.lifting} for further
  explanation.}

\subsection{Overview}
\label{sec:ac.overview}

The traditional approach to defining security can be seen as
\emph{bottom\-/up}. One first defines (at a low level) a computational
model (e.g., a Turing machine or a circuit). Based on this, the
concept of an algorithm for the model and a communication model (e.g.,
based on tapes) are defined. After this, notions of complexity,
efficiency, and finally the security of a cryptosystem can be
defined. The AC framework uses a \emph{top\-/down} approach: in order
to state definitions and develop a theory, one starts from the other
end, the highest possible level of abstraction \--- the composition of
abstract systems \--- and proceeds downwards, introducing in each new
lower level only the minimal necessary specializations. One may give
the analogous example of group theory, which is used to describe
matrix multiplication. In the bottom\-/up approach, one would start
explaining how matrices are multiplied, and then based on this find
properties of the matrix multiplication. In contrast to this, the AC
approach would correspond to first defining the (abstract)
multiplication group and prove theorems already on this level. The
matrix multiplication would then be introduced as a special case of
the multiplicative group.

On a high level of abstraction, a cryptographic protocol, $\pi$, can
be seen as constructing some resource $\aS$ from other resources
$\aR$. For example, information\-/theoretic authentication constructs
an \emph{authentic channel} resource from an \emph{insecure channel}
resource and a \emph{secret key} resource. The resource constructed
can be used by other protocols, e.g., a QKD protocol uses an
\emph{insecure quantum channel} resource and an \emph{authentic
  (classical) channel} to construct a \emph{secret key} resource. In
general however, these resources are not constructed perfectly, there
is some small probability, $\eps$, that the adversary can break the
scheme, e.g., corrupt a message or learn a secret key. If $\pi$
constructs $\aS$ out of $\aR$ within $\eps$, we write
\begin{equation}
  \label{eq:ac} \aR \xrightarrow{\pi,\eps} \aS.
\end{equation}

For a cryptographic construction to be usable
in an arbitrary context, it needs to be composable, i.e., the
following conditions must be fulfilled:
\begin{align*}
\aR \xrightarrow{\pi,\eps} \aS\ \textup{and}\ \aS
  \xrightarrow{\pi',\eps'} \aT & \implies \aR
  \xrightarrow{\pi' \circ \pi,\eps+\eps'} \aT, \\
\aR \xrightarrow{\pi,\eps} \aS \ \textup{and}\  \aR'
  \xrightarrow{\pi',\eps'} \aS' & \implies \aR \| \aR'
  \xrightarrow{\pi|\pi',\eps+\eps'} \aS \| \aS', 
\end{align*}
where $\aR\|\aR'$ is a parallel composition of resources, and $\pi'
\circ \pi$ and $\pi|\pi'$ are sequential and parallel composition of
protocols. In \secref{sec:ac.security} we give a definition for
\eqnref{eq:ac} which satisfies these composable
conditions. Intuitively, the resource $\aR$ along with the protocol
$\pi$ are part of the \emph{real} world, and the resource $\aS$ is the
\emph{ideal} resource we want to build. \eqnref{eq:ac} is then
satisfied if an adversary could, in an ideal world where the ideal
resource is available, achieve anything that she could achieve in the
real world. This argument involves, as a thought experiment, simulator
systems which transform the ideal resource into the real world system
consisting of the real resource and the protocol. But before stating
the security definition, we first define in \secref{sec:ac.resources}
the elements present in \eqnref{eq:ac}, namely resources $\aR$, $\aS$,
a protocol $\pi$, and a pseudo\-/metric\footnoteremember{fn:pm}{A
  function $d : \Omega \times \Omega \to \mathds{R}_+$ is a
  pseudo\-/metric on $\Omega$ if for all $a,b,c \in \Omega$ the three
  following conditions hold: \begin{align*} d(a,a) &= 0, \\ d(a,b) & =
    d(b,a), \\ d(a,b) & \leq d(a,c)+d(c,b).\end{align*} If
  additionally $d(a,b) = 0 \implies a=b$, then $d$ is a metric.} on
the space of resources to quantify the failure $\eps$.

\subsection{Resources, converters and distinguishers}
\label{sec:ac.resources}

On the highest level of abstraction, a \emph{system} is an abstract
object with interfaces that interacts with its environment and with
other systems. Two systems can be composed into a single system by
connecting one interface of each system. In order to define
\eqnref{eq:ac}, it is sufficient to introduce (abstract) systems that
can be composed into larger systems and define an appropriate
pseudo\-/metric. These systems can then be instantiated at a lower
level with any formalism that satisfies the abstract (composition and
metric) properties required by the abstract systems, e.g., random
systems~\cite{Mau02,MPR07} in the classical case, or, if the
underlying computational model is quantum, as a sequence of completely
positive, trace\-/preserving maps with internal memory (e.g., quantum
strategies~\cite{GW07} and combs~\cite{CDP09}). It is however not
necessary to define \--- or even consider \--- these lower levels to
define cryptographic security.

Since the concrete systems used in this work are all very intuitive,
we will not provide a generic (mathematical) definition of these lower
levels. Instead we define each concrete system individually when
needed. 

\paragraph{Resource.} An \emph{$\cI$-resource} is an (abstract) system
with interfaces specified by a set $\cI$ (e.g., $\cI =
\{A,B,E\}$). Each interface $i \in \cI$ is accessible to a user $i$
and provides her or him with certain functionalities. Resources are
equipped with a parallel composition operator, $\|$, that maps two
resources to another resource.

\paragraph{Converter.} To transform one resource into another, we use
\emph{converters}. These are (abstract) systems with two interfaces,
an \emph{inside} interface and an \emph{outside} interface. The inside
interface connects to an interface of a resource, and the outside
interface becomes the new interface of the constructed resource. We
write either $\alpha_i \aR$ or $\aR\alpha_i$ to denote the new
resource with the converter $\alpha_i$ connected at the interface
$i$,\footnote{There is no mathematical difference between $\alpha_i
  \aR$ and $\aR\alpha_i$. It sometimes simplifies the notation to have
  the converters for some players written on the right of the resource
  and the ones for other players on the left, instead of all on the
  same side, hence the two notations.} and $\alpha\aR$ or $\aR\alpha$
for a set of converters $\alpha = \{\alpha_i\}_i$, for which it is
clear to which interface they connect. Converters are equipped with
parallel and sequential composition operators, $|$ and $\circ$, that
map two converters to another converter. A protocol $\pi =
\{\pi_i\}_{i}$ is a set of converters $\pi_i$, indexed by a subset of
interfaces $i \in \cI$.

\paragraph{Example.} To illustrate these notions we model standard
information\-/theoretic authentication. To provide an example it is
necessary to make these systems concrete. We only do this informally
in the following. By using the language of random
systems~\cite{Mau02,MPR07} (or any other appropriate formalism), this
may be made mathematically precise.

\begin{figure}[tb]
\begin{centering}

\begin{tikzpicture}[
      resource/.style={draw,minimum width=3.2cm,minimum height=1cm}]

\small

\node[resource] (keyBox) at (0,0) {};
\node (alice) at (-2.5,0) {Alice};
\node (bob) at (2.5,0) {Bob};
\node (eve) at (0,-1.2) {Eve};
\node[draw] (key) at (0,0) {key};

\draw[sArrow] (key) to node[pos=.55,auto,swap] {$k$} (alice);
\draw[sArrow] (key) to node[pos=.55,auto] {$k$} (bob);

\end{tikzpicture}

\end{centering}
\caption{\label{fig:key.resource}A \emph{secret key resource $\aK$}
  that always gives a key $k$ to Alice and Bob, and nothing to Eve.}
\end{figure}
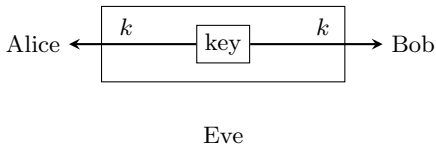

Let us first consider the resources used in the construction of an
authentic channel. A \emph{secret key resource}, $\aK$, can be seen as
a box that outputs a secret key at Alice's and Bob's interfaces, but
does not provide any functionality at Eve's interface. This is
illustrated in \figref{fig:key.resource}.

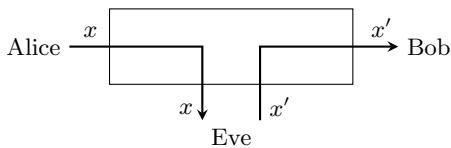
\begin{figure}[tb]
\begin{centering}

\begin{tikzpicture}[
      resource/.style={draw,minimum width=3.2cm,minimum height=1cm}]

\small

\node[resource] (keyBox) at (0,0) {};
\node (alice) at (-2.6,0) {Alice};
\node (bob) at (2.6,0) {Bob};
\node (eve) at (0,-1.2) {Eve};
\node (ajunc) at (eve.north west |- alice) {};
\node (bjunc) at (eve.north east |- bob) {};

\draw[sArrow] (alice) to node[pos=.16,auto] {$x$} (ajunc.center)
to node[pos=.85,auto,swap] {$x$} (eve.north west);
\draw[sArrow] (eve.north east) to node[pos=.18,auto,swap] {$x'$}
(bjunc.center) to node[pos=.88,auto] {$x'$} (bob);

\end{tikzpicture}

\end{centering}
\caption{\label{fig:insecure.resource}An \emph{insecure
    channel $\aC$} from Alice to Bob. Eve obtains Alice's message $x$,
  and can choose what Bob receives, e.g., $x'$.}
\end{figure}

The authentication protocol is run over an \emph{insecure channel},
$\aC$, completely under the control of Eve. One way of modeling such a
channel is to allow Eve to intercept the message sent from Alice to
Bob, and replace it with any message of her choice. This is depicted
in \figref{fig:insecure.resource}.

\begin{figure}[tb]
\begin{centering}

\begin{tikzpicture}[
      resource/.style={draw,minimum width=3.2cm,minimum height=1cm}]

\small

\node[resource] (keyBox) at (0,0) {};
\node (alice) at (-3,0) {Alice};
\node (bob) at (3,0) {Bob};
\node (eve) at (0,-1.2) {Eve};
\node (ajunc) at (eve.north west |- alice) {};

\draw[thick] (alice) to node[pos=.2,auto] {$x$} (0,0) to node[pos=.5]
(ejunc) {} +(160:-.8);
\draw[sArrow] (ajunc.center) to node[pos=.85,auto,swap] {$x$} (eve.north west);
\draw[sArrow] (.8,0) to node[pos=.7,auto] {$x,\bot$} (bob);
\draw[double] (ejunc.center |- eve.north) to node[pos=.15,auto,swap] {$0,1$} (ejunc.center);

\end{tikzpicture}

\end{centering}
\caption{\label{fig:auth.resource}An \emph{authentic channel $\aA$} from
  Alice to Bob. Eve obtains Alice's message $x$, and can choose
  whether Bob receives the original message or an error $\bot$.}
\end{figure}
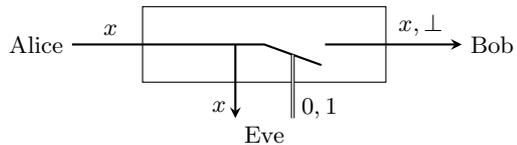

The channel we wish to construct from a secret key and insecure
channel, is an \emph{authentic channel resource}, $\aA$. Ideally, we
would like the channel to always deliver the correct message to the
receiver. This is however impossible to construct from an insecure
channel, since Eve can always jumble the communication between
Alice and Bob. What can be constructed, is a channel which guarantees
that Bob does not receive a corrupted message. He either receives the
correct message sent by Alice, or an error, which symbolizes an
attempt by Eve to change or block the message. This can be modelled by
giving Eve's idealized interface two controls: the first provides her
with Alice's message, the second allows her to decide if Alice's
message should be delivered to Bob or if he gets an error flag
instead. We illustrate this in \figref{fig:auth.resource}. Note that
Eve also has the option of not inputting anything. In a completely
asynchronous model, Bob would then not receiving anything. If the
model allows him to know that a message was sent, he would get an
error flag $\bot$ after the timeout.

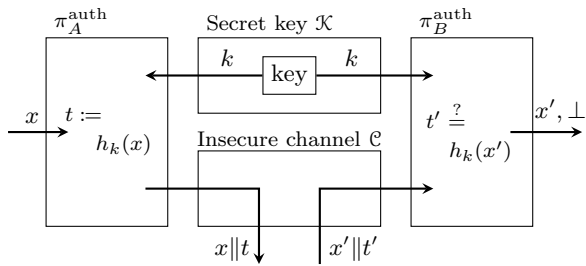
\begin{figure}[tb]
\begin{centering}

\begin{tikzpicture}[
      resource/.style={draw,minimum width=2.4cm,minimum height=1cm},
      protocol/.style={draw,minimum width=1.6cm,minimum height=2.5cm},
      pnode/.style={minimum width=1cm,minimum height=.5cm}]

\small

\def\t{3.85} 
\def\u{2.4} 
\def\v{.75}

\node[pnode] (a1) at (-\u,\v) {};
\node[pnode] (a2) at (-\u,0) {};
\node[pnode] (a3) at (-\u,-\v) {};
\node[protocol,text width=1.1cm] (a) at (-\u,0) {\footnotesize $t
  \coloneqq $\\$\quad \ h_k(x)$};
\node[yshift=-2,above right] at (a.north west) {\footnotesize
  $\pi^{\auth}_A$};
\node[inner sep=0] (alice) at (-\t+.15,0) {};

\node[pnode] (b1) at (\u,\v) {};
\node[pnode] (b2) at (\u,0) {};
\node[pnode] (b3) at (\u,-\v) {};
\node[protocol,text width=1.2cm] (b) at (\u,0) {\footnotesize $t' \stackrel{?}{=}$\\$\quad h_k(x')$};
\node[yshift=-2,above right] at (b.north west) {\footnotesize $\pi^{\auth}_B$};
\node[inner sep=0] (bob) at (\t,0) {};

\node[resource] (keyBox) at (0,\v) {};
\node[draw] (key) at (0,\v) {key};
\node[yshift=-2,above right] at (keyBox.north west) {\footnotesize
  Secret key $\aK$};
\node[resource] (channel) at (0,-\v) {};
\node[yshift=-1.5,above] at (channel.north) {\footnotesize
  Insecure channel $\aC$};
\node (eveleft) at (-.4,-1.75) {};
\node (everight) at (.4,-1.75) {};
\node (ajunc) at (eveleft |- a3) {};
\node (bjunc) at (everight |- b3) {};

\draw[sArrow] (key) to node[auto,swap,pos=.3] {$k$} (a1);
\draw[sArrow] (key) to node[auto,pos=.3] {$k$} (b1);

\draw[sArrow] (alice.center) to  node[auto,pos=.4] {$x$} (a2);
\draw[sArrow] (b2) to node[auto,pos=.7] {$x',\bot$} (bob.center);

\draw[sArrow] (a3) to (ajunc.center)
to node[pos=.8,auto,swap] {$x\|t$} (eveleft.center);
\draw[sArrow] (everight.center) to node[pos=.2,auto,swap] {$x'\|t'$}
(bjunc.center) to (b3);

\end{tikzpicture}

\end{centering}
\caption{\label{fig:auth.real}The \emph{real authentication
    system}. Alice has access to the left interface, Bob to the right
  interface and Eve to the lower interface. The converters
  $(\pi^{\auth}_A,\pi^{\auth}_B)$ of the authentication protocol are
  connected to the resource $\aK \| \aC$ consisting of a secret key
  and an insecure channel. $\pi^{\auth}_A$ generates a tag $t$ for the
  message $x$ using the secret key $k$. $\pi^{\auth}_B$ checks if the
  received message and tag, $x'\|t'$, match.}
\end{figure}

For Alice to send an authenticated message to Bob, her part of the
protocol, $\pi^{\auth}_A$, is a converter, which gets a key $k$ at the
inner interface from an ideal key resource, a message $x$ at the outer
interface from Alice, computes a tag $t = h_k(x)$ and sends the
concatenation of the message and tag, $x\|t$ through its inside
interface down the insecure channel. Bob's part of the protocol,
$\pi^{\auth}_B$, gets the same key as Alice, $k$, from the ideal key
resource at its inner interface, a message $x'\|t'$ from the channel
at its inner interface, and outputs at its outer interface either $x'$
if $t' = h_k(x')$, or an error $\bot$ otherwise. This is depicted in
\figref{fig:auth.real}.

\paragraph{Distinguisher.} To measure how close two resources are, we
define a pseudo\-/metric\footnote{Recall \footnoteref{fn:pm} on
  \pref{fn:pm}.} on the space of resources. We do this with the help
of a \emph{distinguisher}. For $n$\=/interface resources a
distinguisher $\aD$ is a system with $n+1$ interfaces, where $n$
interfaces connect to the interfaces of a resource $\aR$ and the other
(outside) interface outputs a bit. For a class of distinguishers
$\mathds{D}$, the induced pseudo\-/metric, the distinguishing
advantage, is \[d(\aR,\aS) \coloneqq \max_{\aD \in \mathds{D}}
\Pr\left[\aD\aR = 1\right] - \Pr\left[\aD\aS = 1\right],\] where
$\aD\aR$ is the binary random variable corresponding to $\aD$
connected to $\aR$. In this work we study information\-/theoretic
security, and therefore the only class of distinguishers that we
consider is the set of all distinguishers. If $d(\aR,\aS) \leq \eps$,
we say that the two resources are $\eps$-close and sometimes write
$\aR \close{\eps} \aS$; or $\aR = \aS$ if $\eps = 0$.

In \secref{sec:ac.dist} we introduce some of the properties of the
distinguishing advantage. We also show how it can be related to the
statistical distance between the probability distributions of the
underlying real and ideal systems.

\begin{rem}
  \label{rem:ac.lifting}
  If two \emph{classical} systems $\aR$ and $\aS$ are
  indistinguishable for the set of all classical distinguishers, then
  they are also indistinguishable for the set of all quantum
  distinguishers, since classical computers (or distinguishers) can
  simulate quantum ones. A secure classical protocol is then secure
  regardless of whether it is later composed with quantum
  systems. This is immediate from the framework without needing to
  define quantum security.\footnote{In the case of computational
    security, the class of classical distinguishers efficiently
    implementable on a quantum computer guarantees security under
    composition with quantum systems.}
\end{rem}

\subsection{Security definition}
\label{sec:ac.security}

We can now define the security of a protocol in the three party
setting with honest Alice and Bob, and dishonest Eve.
\begin{deff}[Composable security \cite{MR11}]
\label{def:ac.security}
Let $\aR$ and $\aS$ be resources with interfaces $\cI = \{A,B,E\}$. We
say that a protocol $\pi = (\pi_A,\pi_B)$ (securely)
constructs $\aS$ out of $\aR$ within $\eps$, and write
$\aR \xrightarrow{\pi,\eps} \aS$, if the two following
conditions hold:
\begin{enumerate}[label=\roman*), ref=\roman*]
\item \label{eq:def.cor} For converters $\sharp_E$ and $\flat_E$ which
  emulate an honest behavior at Eve's interfaces,
  \[d(\pi\aR\sharp_E,\aS\flat_E) \leq \eps.\]
\item \label{eq:def.sec} There exists a converter $\sigma_E$\--- which
  we call simulator \--- such that
  \[  d(\pi\aR,\aS\sigma_E) \leq \eps.\]
\end{enumerate}
If it is clear from the context what resources $\aR$ and $\aS$ are
meant, we simply say that $\pi$ is $\eps$\=/secure.
\end{deff}

The first of these two conditions can be seen as capturing correctness
of the protocol in the case where no adversary is present, i.e., when
the distinguisher does not access the adversarial controls covered by
$\sharp_E$ and $\flat_E$. If however the opponent is active, the
converters $\sharp_E,\flat_E$ are removed and the distinguisher has
full access to Eve's interfaces. This is captured by the second
condition in \defref{def:ac.security}. In the case of authentication,
it is not hard to see that $\pi^{\auth}(\aK\|\aC)\sharp_E =
\aA\flat_E$. If the converter $\sharp_E$ faithfully transmits the
string $x \| t$ to Bob, the system $\pi^{\auth}(\aK\|\aC)\sharp_E$ is
a simple channel that transmits Alice's message $x$ to Bob. If the
converter $\flat_E$ at Eve's interface of the ideal authentic channel
$\aA$ inputs the bit allowing Alice's message through, we then also
have a channel that transmits Alice's message $x$ to Bob.

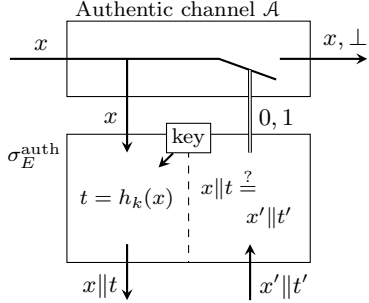
\begin{figure}[tb]
\begin{centering}

\begin{tikzpicture}[
      resource/.style={draw,minimum width=3.2cm,minimum height=1cm},
      simulator/.style={draw,minimum width=3.2cm,minimum height=1.7cm},
      snode/.style={minimum width=1.1cm,minimum height=1.2cm}]

\small

\def\t{2.35} 
\def\u{-1.1}
\def\v{.75}
\def\w{-2.45} 

\node[resource] (channel) at (0,\v) {};
\node[yshift=-1.5,above right] at (channel.north west) {\footnotesize
  Authentic channel $\aA$};
\node (alice) at (-\t,\v) {};
\node (bob) at (\t,\v) {};

\node[simulator] (sim) at (0,\u) {};
\node[xshift=1.5,below left] at (sim.north west) {\footnotesize
  $\sigma^{\auth}_E$};
\node[snode,ellipse] (sleft) at (-.809,\u) {};
\node[snode] (sright) at (.809,\u) {};
\draw[dashed] (sim.north) to (sim.south);

\node (ajunc) at (sleft |- alice) {};
\node (bjunc) at (sright |- bob) {};

\draw[thick] (alice.center) to node[pos=.15,auto] {$x$} (.4,\v) to node[pos=.54] (ejunc) {} +(160:-.8);
\draw[sArrow] (ajunc.center) to node[pos=.63,auto,swap] {$x$} (sleft);
\draw[sArrow] (1.2,\v) to node[pos=.75,auto] {$x,\bot$} (bob.center);
\draw[double] (sright) to node[pos=.4,auto,swap] {$0,1$} (ejunc.center);

\node (sltext) at (-.809,\u) {\footnotesize $t = h_k(x)$};
\node[text width=1.3cm] (srtext) at (.809,\u) {\footnotesize $x\|t
  \stackrel{?}{=}$\\$\quad \quad x'\|t'$};

\node (eveleft) at (sleft |- 0,\w) {};
\node (everight) at (sright |- 0,\w) {};
\draw[sArrow] (sleft) to node[pos=.75,auto,swap] {$x\|t$} (eveleft.center);
\draw[sArrow] (everight.center) to node[pos=.25,auto,swap] {$x'\|t'$}
(sright);

\node[draw,inner sep=2,fill=white] (key) at (0,\u+.8) {\footnotesize key};
\draw[sArrow] (key) to (sleft);

\end{tikzpicture}

\end{centering}
\caption{\label{fig:auth.ideal}The \emph{ideal authentication
    system}. Alice has access to the left interface, Bob to the right
  interface and Eve to the lower interface. The simulator
  $\sigma^{\auth}_E$ is plugged into the $E$\=/interface of the ideal
  authentic channel $\aA$. It generates its own tag $t = h_k(x)$ to
  simulate the message $x\|t$ on the insecure channel, and notifies
  the ideal authentic channel to output an error $\bot$ if any part of
  $x\|t$ got changed.}
\end{figure}

To show that condition \eqref{eq:def.sec} is fulfilled (for some well
chosen family of hash functions), we need to find a simulator
$\sigma^{\auth}_E$ that can recreate the real $E$\=/interface while
accessing just the idealized one. An obvious choice for the simulator
is to first generate its own key $k$ and output $x\|h_k(x)$. Then upon
receiving $x'\|t'$, it checks if $x'\|t' = x\|h_k(x)$ and sets the
control bit of the ideal authentic channel $\aA$ accordingly. If the
distinguisher chooses to provide the inputs in reversed order, first
input $x'\|t'$ at the $E$\=/interface and then choose a message $x$
that it inputs at the $A$\=/interface, the simulator always sets the
control bit on the authentic channel to output an error $\bot$ at the
$B$\=/interface and does not choose the key $k$ uniformly at random
from the entire set, but only from the subset of keys such that
$h_k(x') \neq t'$. We illustrate this simulator in
\figref{fig:auth.ideal}.

\subsection{Distinguishing advantage}
\label{sec:ac.dist}

Because it is a pseudo\-/metric, the distinguishing advantage respects
the triangle inequality. For any resources $\aR,\aS,\aT$,
 \begin{equation}
    \label{eq:dist.tri}
    d(\aR,\aS) \leq d(\aR,\aT) + d(\aT,\aS).
\end{equation}

The distinguishing advantage is also non\-/increasing under
composition with any other (abstract) system. For any systems
$\aR,\aS,\aT$,
\begin{equation}
  \label{eq:dist.comp}
  d(\aR\aT,\aS\aT) \leq d(\aR,\aS).
\end{equation}
This holds because the distinguisher can run $\aT$ internally, so the
maximization over all distinguishers already includes the composition
with $\aT$.\footnote{This also holds in the case of computational
  security, since the class of efficiently implementable
  distinguishers is closed under composition with efficiently
  implementable systems $\aT$.}

Resources can be seen as interactive black boxes. Each interface has
various ports which either take inputs or produce outputs. A
distinguisher holding a resource can input the value of its choosing
at any port (as long as the value is from the permitted alphabet for
that port), and store the outputs it receives. For example, if it
holds the real authentication system of \figref{fig:auth.real}, it can
input some message $x$ at the $A$\=/interface and receives $x\|t$ at
the $E$\=/interface. This in- and output pair is described by a joint
probability distribution $P_{XT}$ where $X$ is the label of the input
port on the $A$\=/interface, and $T$ denotes the port outputting $t$
on the $E$\=/interface.\footnote{$x$ is also output at the
  $E$\=/interface, but since it is always identical to the input at
  port $X$, we avoid writing it a second time.} After having
interacted with all the ports, the distinguisher can make a guess as
to whether it is holding the real or the ideal system. Its
distinguishing advantage is given by the statistical distance
(\eqnref{eq:sd}) between the probability distributions of the
corresponding in- and outputs. By maximizing the statistical distance
over all permutations of the input ports and all choices of inputs we
get the distinguishing advantage between the two resources.

For example, the real and ideal authentication systems of
Figures~\ref{fig:auth.real} and \ref{fig:auth.ideal} take two inputs,
a message $x$ at the $X$ port of the $A$\=/interface and a string
$x'\|t'$ at the $X'T'$ port of the $E$\=/interface, so there are also
two orderings in which the distinguisher can interact with the
system. In the standard authentication literature these two possible
orders are referred to as \emph{substitution attack} (when the
distinguisher first obtains $x\|t$ and modifies it to create $x'\|t'$)
and \emph{impersonation attack} (when the distinguisher directly
generates $x'\|t'$ without having seen a valid message\-/tag pair).

In the case of a substitution attack, let the distinguisher choose
some input $x$ with probability $P_X(x)$. It then receives $x\|t$
where $P_{T|X}(t|x)$ is defined by the scheme used, and is identical
in the real and ideal situations. It chooses some $x'\|t'$ with
probability $P_{X'T'|XT}(x',t'|x,t)$, and receives a final output $y$
which can take two values, $x'$ or $\bot$. The probability of taking
either of these values is however different in the real and ideal
settings. In the real one, $y = x'$ if $h_k(x') = t'$, and in the
ideal one, $y = x'$ if $x'\|t'=x\|t$. Let $P_{Y|XTX'T'}$ and
$Q_{Y|XTX'T'}$ be the probability distributions of $y$ in these two
settings. The real world is then completely described by $P_{XTX'T'Y}$
and the ideal world by $Q_{XTX'T'Y}$, where $Q_{XTX'T'} =
P_{XTX'T'}$. The distinguisher can now pick a subset of values $\cD
\subseteq \cX \times \cT \times \cX \times \cT \times (\cX \cup
\{\bot\})$, for which it will guess that it is holding the real
system, i.e., for all $(x,t,x',t',y) \in \cD$ it outputs $1$ and for
all others it outputs $0$. Its advantage is then
\begin{multline}
  \max_{\cD} \sum_{(x,t,x',t',y) \in \cD}
  \left(P_{XTX'T'Y}(x,t,x',t',y) \right. \\ \shoveright{\left. -\> Q_{XTX'T'Y}(x,t,x',t',y) \right)} \\
  \shoveleft{\quad = \frac{1}{2}
    \sum_{x,t,x',t',y} \left| P_{XTX'T'Y}(x,t,x',t',y) \right.} \\
  \left. -\> Q_{XTX'T'Y}(x,t,x',t',y)
  \right|, \label{eq:auth.sub} \end{multline} the statistical distance
between $P_{XTX'T'Y}$ and $Q_{XTX'T'Y}$.

In the case of an impersonation attack, the distinguisher will gain
nothing from inputting $x$, after having input $x'\|t'$: if it was
lucky and the real system accepted $x'\|t'$, it can already
distinguish them with advantage $1$, and if $x'\|t'$ was rejected,
both systems will generate a $x\|t$ with the same distribution since
the simulator picks $k$ from all keys such that $h_k(x') \neq t'$. We
can thus consider only the ports $X'T'$ and $Y$. The advantage is then
\begin{equation}
  \label{eq:auth.imp} \frac{1}{2}
  \sum_{x',t',y} \left| P_{X'T'Y}(x',t',y) -
  Q_{X'T'Y}(x',t',y) \right|.
\end{equation}

In \appendixref{app:stdauth} we provide a proof that
\eqnsref{eq:auth.sub} and \eqref{eq:auth.imp} are both bounded by
$\eps$ for all $P_{X},P_{X'T'|XT}$ and $P_{X'T'}$, respectively, if
$\eps$\=/ASU$_2$ hash functions are used by the protocol.

\section{Arbitrary key recycling with synchronization}
\label{sec:sync}

The authentication protocol depicted in \figref{fig:auth.real}
requires a new hash function and therefore a (completely) new secret
key $k$ for every new message. We study in this section a scheme which
allows part of the secret key to be reused. Intuitively, this is
possible because the recycled part of the key is almost uniform
conditioned on all the inputs, outputs and the adversary's information
produced throughout the protocol.

It is however vital that the key be recycled only \emph{after} the
receiver, Bob, either obtains the message or decides not to accept
anymore an incoming message. Otherwise Eve could first obtain the
recycled key from whatever protocol uses it next and then modify the
authenticated message. In this section we thus assume a mild form of
synchronization between sender and receiver: an incoming message will
not be accepted by Bob after the key has been recycled by Alice.

For practicality we model this as a timeout, i.e., after a predefined
amount of time has passed the sender can recycle the key and the
receiver will not accept a message arriving late. It can however be
achieved by other means, e.g., Bob sends an authenticated
acknowledgment of receptions. This occurs naturally if the messages
alternate, one from Alice to Bob and the next from Bob to Alice. If
Alice successfully authenticates an incoming message, this guarantees
\--- up to an error $\eps$ \--- that her previous message was received
by Bob, who otherwise would not have sent the next message. So she can
already recycle the key, even if the timeout has not
occurred. Likewise Bob can recycle the key used for the messages in
the other direction as soon as he receives an authenticated response
from Alice.

\subsection{Ideal resource}
\label{sec:sync.ideal}

\begin{figure}[tb]
\begin{centering}

\begin{tikzpicture}[
      resource/.style={draw,minimum width=3.2cm,minimum height=1.8cm}]

\small

\def\t{2.55}
\def\w{.6}
\def\b{.4}

\def\v{.35}
\def\e{-1.3}

\node[resource] (keyBox) at (0,0) {};
\node[yshift=-1.5,above right] at (keyBox.north west) {\footnotesize
  Authentic channel $\aA^+$};
\node[draw] (key) at (0,\v) {key};
\node (a1) at (-\t,\v) {};
\node (a2) at (-\t,-\v) {};
\node (b1) at (\t,\v) {};
\node (b2) at (\t,-\v) {};
\node (e2) at (-\w,\e) {};
\node (e3) at (\w,\e) {};
\node (leak) at (-\w,-\v) {};
\node (s2) at (\w,-\v) {};

\draw[thick] (a2.center) to node[pos=.2,auto] {$x$} (\w-\b,-\v) to node[pos=.53]
(handle3) {} +(340:2*\b);
\draw[sArrow] (\w+\b,-\v) to node[pos=.7,auto] {$x,\bot$} (b2.center);
\draw[sArrow] (key) to node[pos=.75,auto] {$k$} (b1.center);
\draw[sArrow] (key) to node[pos=.75,auto,swap] {$k$} (a1.center);
\draw[sArrow] (leak.center) to node[pos=.8,auto] {$x$} (e2.center);
\draw[double] (e3.center) to node[pos=.15,auto,swap] {$0,1$} (handle3.center);

\end{tikzpicture}

\end{centering}
\caption{\label{fig:recycle.resource}An \emph{authentic channel with
    key recycling $\aA^+$}. Alice has access to the left interface,
  Bob to the right interface and Eve to the lower interface. Eve
  obtains Alice's message $x$, and can choose whether Bob receives the
  original message or an error $\bot$. After providing Bob with the
  error or message, the resource generates a new key $k$, which is
  given to both Alice and Bob.}
\end{figure}
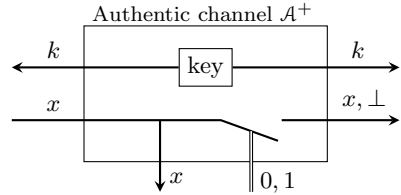

An authentic channel with key recycling, $\aA^+$, does not only
guarantee that the message delivered has not been tampered with, it
also provides the users with a new (recycled) key that they can use
for any arbitrary application requiring a shared secret key. It can be
seen as the combination of an authentic channel
(\figref{fig:auth.resource}) and a secret key resource
(\figref{fig:key.resource}), which we illustrate in
\figref{fig:recycle.resource}.

This ideal resource must follow the same timing rules as the real
system. As Bob can recycle the key has soon as he receives the message
from Alice, the resource $\aA^+$ outputs the new key $k$ at the
$B$\=/interface at the same time as the message $x$ or the decision to
reject it, $\bot$. However, the new key $k$ is only output at the
$A$\=/interface when the timeout occurs. In the real system, Bob's protocol can
output an error $\bot$ if no message has arrived when the time is up,
since no late message is accepted. The ideal resource $\aA^+$ thus
also outputs $\bot$ in this case.

\subsection{Protocol}
\label{sec:sync.protocol}

Like for standard authentication, we construct the ideal resource from
a secret key and an insecure channel. As stated in the introduction,
the main idea is to encrypt the tag $t$ appended to the message in
standard authentication with a OTP, and then reuse the same hash
function. For this, Alice's protocol $\pi^{\auth+}_A$ splits the
shared secret key in two parts $k = k_1\|k_2$, and uses them to
generate a tag $t \coloneqq h_{k_1}(x) \xor k_2$ for the message $x$,
where $\{h_k\}_k$ is a family of $\eps$\=/AXU hash functions. The
string $x\|t$ is sent to Bob on the insecure channel $\aC$. After a
predefined amount of time has elapsed, Alice outputs the key $k_1$ for
recycling.

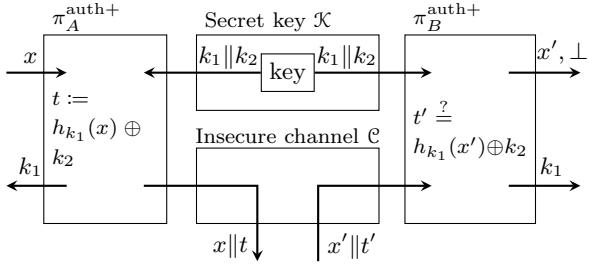
\begin{figure}[tb]
\begin{centering}

\begin{tikzpicture}[
      resource/.style={draw,minimum width=2.4cm,minimum height=1cm},
      protocol/.style={draw,minimum width=1.6cm,minimum height=2.5cm},
      pnode/.style={minimum width=1cm,minimum height=.5cm}]

\small

\def\t{3.85} 
\def\u{2.4} 
\def\v{.75}

\node[pnode] (a1) at (-\u,\v) {};
\node[pnode] (a2) at (-\u,0) {};
\node[pnode] (a3) at (-\u,-\v) {};
\node[protocol,text width=1.4cm] (a) at (-\u,0) {\footnotesize $t
  \coloneqq $\\$h_{k_1}(x) \xor k_2$};
\node[yshift=-2,above right] at (a.north west) {\footnotesize
  $\pi^{\auth+}_A$};
\node[inner sep=0] (alice1) at (-\t+.15,\v) {};
\node[inner sep=0] (alice2) at (-\t+.15,-\v) {};

\node[pnode] (b1) at (\u,\v) {};
\node[pnode] (b2) at (\u,0) {};
\node[pnode] (b3) at (\u,-\v) {};
\node[protocol,text width=1.5cm] (b) at (\u,0) {\footnotesize $t'
  \stackrel{?}{=}$\\$h_{k_1}(x') \xor k_2$};
\node[yshift=-2,above right] at (b.north west) {\footnotesize $\pi^{\auth+}_B$};
\node[inner sep=0] (bob1) at (\t,\v) {};
\node[inner sep=0] (bob2) at (\t,-\v) {};

\node[resource] (keyBox) at (0,\v) {};
\node[draw] (key) at (0,\v) {key};
\node[yshift=-2,above right] at (keyBox.north west) {\footnotesize
  Secret key $\aK$};
\node[resource] (channel) at (0,-\v) {};
\node[yshift=-1.5,above] at (channel.north) {\footnotesize
  Insecure channel $\aC$};
\node (eveleft) at (-.4,-1.75) {};
\node (everight) at (.4,-1.75) {};
\node (ajunc) at (eveleft |- a3) {};
\node (bjunc) at (everight |- b3) {};

\draw[sArrow] (key) to node[auto,swap,pos=.275,yshift=-2] {$k_1\|k_2$} (a1);
\draw[sArrow] (key) to node[auto,pos=.275,yshift=-2] {$k_1\|k_2$} (b1);

\draw[sArrow] (alice1.center) to  node[auto,pos=.4] {$x$} (a1);
\draw[sArrow] (b1) to node[auto,pos=.75] {$x',\bot$} (bob1.center);

\draw[sArrow] (a3) to  node[auto,pos=.6,swap] {$k_1$} (alice2.center);
\draw[sArrow] (b3) to node[auto,pos=.6] {$k_1$} (bob2.center);

\draw[sArrow] (a3) to (ajunc.center)
to node[pos=.8,auto,swap] {$x\|t$} (eveleft.center);
\draw[sArrow] (everight.center) to node[pos=.2,auto,swap] {$x'\|t'$}
(bjunc.center) to (b3);

\end{tikzpicture}

\end{centering}
\caption{\label{fig:recycle.real}The \emph{real authentication with
    key recycling} system. Alice has access to the left interface, Bob
  to the right interface and Eve to the lower
  interface. $\pi^{\auth+}_A$ first receives a message $x$ and key
  $k_1\|k_2$, and sends $x\|t$ to Bob. $\pi^{\auth+}_B$ checks whether
  $t' = h_{k_1}(x') \xor k_2$ and outputs the result $x'$ or
  $\bot$. Both converters recycle $k_1$, Bob's immediately and Alice's
  after the timeout.}
\end{figure}

Bob's protocol $\pi^{\auth+}_B$, upon receiving $x'\|t'$ on the
insecure channel, checks whether $t' = h_{k_1}(x') \xor k_2$, and if
so, it accepts and outputs the message $x'$ at its outer
interface. Otherwise, it outputs an error symbol $\bot$. If no message
has been received before the timeout, it outputs an error and does not
accept incoming messages any more. The key $k_1$ is also output for
recycling as soon as accepted or rejected, or a timeout occurred. The
entire protocol is depicted in \figref{fig:recycle.real}.

If Eve does not intervene, security is modeled by placing a converter
$\sharp_E$ over the $E$\=/interface of the system, which connects the
out- and in-ports of the insecure channel and emulates an honest
behavior by forwarding $x\|t$ to Bob.

\subsection{Security}
\label{sec:sync.security}

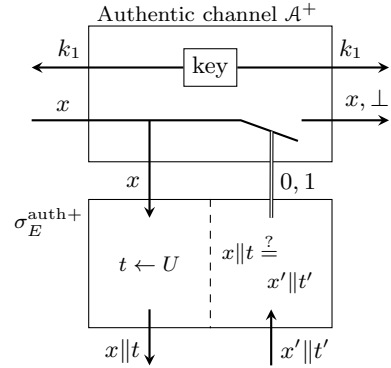
\begin{figure}[tb]
\begin{centering}

\begin{tikzpicture}[
      resource/.style={draw,minimum width=3.2cm,minimum height=1.8cm},
      simulator/.style={draw,minimum width=3.2cm,minimum height=1.7cm},
      snode/.style={minimum width=1.1cm,minimum height=1.2cm}]

\small

\def\t{2.35} 
\def\u{-1.1}
\def\v{1.15}
\def\vv{.35}
\def\w{-2.45} 

\node[resource] (channel) at (0,\v) {};
\node[draw] (key) at (0,\v+\vv) {key};
\node[yshift=-1.5,above right] at (channel.north west) {\footnotesize
  Authentic channel $\aA^+$};
\node (alice1) at (-\t,\v+\vv) {};
\node (bob1) at (\t,\v+\vv) {};
\node (alice2) at (-\t,\v-\vv) {};
\node (bob2) at (\t,\v-\vv) {};

\node[simulator] (sim) at (0,\u) {};
\node[xshift=1.5,below left] at (sim.north west) {\footnotesize
  $\sigma^{\auth+}_E$};
\node[snode] (sleft) at (-.8,\u) {};
\node[snode] (sright) at (.8,\u) {};
\draw[dashed] (sim.north) to (sim.south);

\node (ajunc) at (sleft |- alice2) {};
\node (bjunc) at (sright |- bob2) {};

\draw[sArrow] (key) to node[pos=.75,auto] {$k_1$} (bob1.center);
\draw[sArrow] (key) to node[pos=.75,auto,swap] {$k_1$} (alice1.center);

\draw[thick] (alice2.center) to node[pos=.15,auto] {$x$} (.4,\v-\vv) to node[pos=.54] (ejunc) {} +(160:-.8);
\draw[sArrow] (ajunc.center) to node[pos=.63,auto,swap] {$x$} (sleft);
\draw[sArrow] (1.2,\v-\vv) to node[pos=.75,auto] {$x,\bot$} (bob2.center);
\draw[double] (sright) to node[pos=.4,auto,swap] {$0,1$} (ejunc.center);

\node (sltext) at (-.8,\u) {\footnotesize $t \leftarrow U$};
\node[text width=1.3cm] (srtext) at (.8,\u) {\footnotesize $x\|t
  \stackrel{?}{=}$\\$\quad \quad x'\|t'$};

\node (eveleft) at (sleft |- 0,\w) {};
\node (everight) at (sright |- 0,\w) {};
\draw[sArrow] (sleft) to node[pos=.75,auto,swap] {$x\|t$} (eveleft.center);
\draw[sArrow] (everight.center) to node[pos=.25,auto,swap] {$x'\|t'$}
(sright);


\end{tikzpicture}

\end{centering}
\caption{\label{fig:recycle.ideal}The \emph{ideal authentication with
    key recycling} system. Alice has access to the left interface, Bob
  to the right interface and Eve to the lower interface. Upon
  receiving a message $x$ from the ideal channel, the simulator
  $\sigma^{\auth+}_E$ picks a tag $t$ uniformly at random and outputs
  $x\|t$ at the $E$\=/interface. When it receives $x'\|t'$, it checks
  if this is equal to $x\|t$, and sets the control bit on the ideal
  channel accordingly.}
\end{figure}

To prove that the protocol depicted in \figref{fig:recycle.real}
constructs the ideal resource of \figref{fig:recycle.resource}, we
need to find a simulator $\sigma^{\auth+}_E$ that when plugged into
the idealized $E$\=/interface of the authentication channel with key
recycling recreates the real interface of
\figref{fig:recycle.real}. We do this with the simulator from
\figref{fig:recycle.ideal}.

Similar to that of \figref{fig:auth.ideal}, the simulator
$\sigma^{\auth+}_E$ generates a tag $t$ uniformly at random, which it
outputs with the message $x$. Then upon receiving $x'\|t'$, it checks
if the message has been changed and notifies the ideal resource
$\aA^+$ of this. In the case of an impersonation attack, in which
first $x'\|t'$ is input at the $E$\=/interface and then $x$ at the
$A$\=/interface, the simulator still generates the tag $t$ uniformly
at random.

\begin{thm}
  \label{thm:auth}
  The protocol $(\pi^{\auth+}_A,\pi^{\auth+}_B)$
  described above is $\eps$\=/secure, i.e., \[\aK\|\aC
  \xrightarrow{\pi^{\auth+},\eps} \aA^+.\]
\end{thm}

As described in \secref{sec:ac.dist}, the real and ideal systems
depicted in Figures~\ref{fig:recycle.real} and \ref{fig:recycle.ideal}
can be seen as interactive black boxes. The distinguisher is given one
of the two, can input the values of its choosing, and based on the
outputs, it must guess which of the two it is holding. Similar to the
discussion in \secref{sec:ac.dist}, for every ordering of the inputs
and every (probabilistic) choice of inputs we derive a probability
distribution of the in- and outputs for the real and ideal systems,
and need to bound the statistical distance between the two. To do
this, we show that the probability of an event which includes a
rejected corrupted message is always larger in the ideal case. Since
the probability of an event which includes an accepted corrupted
message is always larger in the real case (it never happens in the
ideal case), we have from \eqnref{eq:sd} that the statistical distance
reduces to the difference between the distributions in the case of
accepted corrupted messages. But this is simply the probability of
accepting a corrupted message in the real setting.

\begin{proof}
In the case where no adversary is present, we trivially have
$$\pi^{\auth+}(\aK\|\aC)\sharp_E= \aA^+ \flat_E,$$ since both systems are
equivalent to a channel that faithfully transmits a message from Alice
to Bob. So we only need to analyze the case of an active adversary.

Each box in Figures~\ref{fig:recycle.real} and \ref{fig:recycle.ideal}
takes two inputs \--- a message $x$ at the $A$\=/interface, and
possibly corrupted message $x'||t'$ at the $E$\=/interface. So the
distinguisher can choose to start by first providing either some $x$
or some $x'||t'$ to the system it is holding.

We first analyze the latter, the impersonation attack. Note that after
choosing the input $x'\|t'$ and receiving Bob's output and the
recycled key at the $B$\=/interface, the distinguisher can still input
a value $x$ at the $A$\=/interface and receive the corresponding tag
$t$. Let $P_{X'T'YKXT}$ and $Q_{X'T'YKXT}$ describe the distributions
of the real and ideal systems. We show in the following that for all
$x',t',k_1,x,t$,
\begin{multline}
  \label{eq:thm.if}
  P_{X'T'YKXT}(x',t',\bot,k_1,x,t) \\ \leq Q_{X'T'YKXT}(x',t',\bot,k_1,x,t).
\end{multline}
Since $Q_{X'T'YKXT}(x',t',x',k_1,x,t) = 0$, the statistical distance
is then given by
\begin{multline*}
\sum_{x',t',k_1,x,t} P_{X'T'YKXT}(x',t',x',k_1,x,t) \\
\begin{aligned} & = \sum_{x',t'} P_{X'T'Y}(x',t',x') \\ 
& = \sum_{x',t'} P_{X'T'}(x',t') P_{Y|X'T'}(x'|x',t')\\
& = \sum_{x',t'}  P_{X'T'}(x',t') 2^{-|k_2|} = 2^{-|k_2|} ,
\end{aligned}\end{multline*}
where to reach the last line we used \eqnref{eq:xorhash1},
namely \begin{equation*}
  P_{Y|X'T'}(x'|x',t') = \Pr_{k_1,k_2} \left[ h_{k_1}(x') \xor k_2 =
    t' \right] = \frac{1}{2^{|k_2|}}.\end{equation*}

It now remains to show that \eqnref{eq:thm.if} holds. In the ideal case
the tag $t$ and key $k_1$ are always generated uniformly at random and
$\bot$ is output of the $Y$ port with probability $1$. So
\begin{align*}& Q_{X'T'YKXT}(x',t',\bot,k_1,x,t) \\
  & = Q_{X'T'}(x',t') Q_{Y|X',T'}(\bot|x',t') \\ & \quad \ 
  Q_{K|X'T'Y}(k_1|x',t',\bot) Q_{X|X'T'YK}(x|x',t',\bot,k_1) \\ &
  \quad \ Q_{T|X'T'YKX}(t|x',t',\bot,k_1,x) \\ & = P_{X'T'}(x',t')
  P_{X|X'T'YK}(x|x',t',\bot,k_1) 2^{-|k_1|-|k_2|}. \end{align*} In the
real case, the probability of accepting $x'$ is exactly $2^{-|k_2|}$,
so the probability of detecting the impersonation is \[
P_{Y|X',T'}(\bot|x',t') = 1 - 2^{-|k_2|}.\] As shown in
\eqnref{eq:xorhash1}, the probability of detecting the impersonation
is actually independent of the value of the recycled key,
hence \[P_{K|X'T'Y}(k_1|x',t',\bot) = 2^{-|k_1|}.\] Finally, knowledge
of an invalid pair $(x',t')$ and the recycled key $k_1$ excludes
exactly one possible value for $t$, the others are all equally
likely. So
\[P_{T|X'T'YKX}(t|x',t',\bot,k_1,x) \leq \frac{1}{2^{|k_2|}-1}.\]
Putting this together, \begin{align*}& P_{X'T'YKXT}(x',t',\bot,k_1,x,t) \\
  & = P_{X'T'}(x',t') P_{Y|X',T'}(\bot|x',t') \\ & \quad \ 
  P_{K|X'T'Y}(k_1|x',t',\bot) P_{X|X'T'YK}(x|x',t',\bot,k_1) \\ &
  \quad \  P_{T|X'T'YKX}(t|x',t',\bot,k_1,x) \\ & \leq
  P_{X'T'}(x',t') P_{X|X'T'YK}(x|x',t',\bot,k_1) 2^{-|k_1|-|k_2|} \\ &
  = Q_{X'T'YKXT}(x',t',\bot,k_1,x,t). \end{align*}

We now consider the substitution attack. In both the real and ideal
cases, the distinguisher chooses some $x$, obtains $x\|t$ and then has
to pick some $x'\|t'$. Note that if the distinguisher chooses $x' =
x$, both systems behave identically: they reject $x'$ if $t' \neq t$,
accept it if $t' = t$, and output a recycled key which is uniform and
independent from $(x,t,x',t')$ (see \eqnref{eq:xorhash1} for the
independence of $k_1$). So it is sufficient to consider strategies for
which $x' \neq x$. Thus, w.l.o.g.\ we assume that the distinguisher
chooses values for the inputs at ports $X$ and $X'$ such that for all
$x$, $P_{XX'}(x,x) = Q_{XX'}(x,x) = 0$. In the following, we show that
for all $x,t,x',t',k_1$
\begin{multline}
  \label{eq:thm.iff}
  P_{XTX'T'YK}(x,t,x',t',\bot,k_1) \\ \leq Q_{XTX'T'YK}(x,t,x',t',\bot,k_1).
\end{multline}
Since $Q_{XTX'T'YK}(x,t,x',t',x',k_1) = 0$, the statistical distance
is then given by
\begin{multline*}
\sum_{x,t,x',t',k_1} P_{XTX'T'YK}(x,t,x',t',x',k_1) \\
\begin{aligned} & = \sum_{x,t,x',t'} P_{XTX'T'Y}(x,t,x',t',x') \\ 
& = \sum_{x,t,x',t'} P_{XTX'T'}(x,t,x',t')
\\ & \qquad \qquad  \qquad \quad P_{Y|XTX'T'}(x'|x,t,x',t')\\
& \leq \sum_{x,t,x',t'}  P_{XTX'T'}(x,t,x',t') \eps = \eps,
\end{aligned}\end{multline*}
where to reach the last line we used \begin{multline*}
  P_{Y|XTX'T'}(x'|x,t,x',t') = \\
  \Pr_{k_1,k_2} \left[ h_{k_1}(x') \xor k_2 = t' \middle| h_{k_1}(x)
    \xor k_2 = t \right]\end{multline*} and \eqnref{eq:xorhashotp}.

It now remains to prove that \eqnref{eq:thm.iff} holds. Note
that \begin{multline*}Q_{XTX'T'YK}(x,t,x',t',\bot,k_1) \\ =
  P_{XTX'T'}(x,t,x',t')2^{-|k_1|},\end{multline*} so it is sufficient
to show that for all $x,t,x',t',k_1$, $$P_{YK|XTX'T'}(\bot,k_1|x,t,x',t')
\leq 2^{-|k_1|}.$$ This is however immediate from \begin{multline*}
  P_{YK|XTX'T'}(\bot,k_1|x,t,x',t') \\ \begin{aligned} &  =
  P_{K|XTX'T'}(k_1|x,t,x',t') \\ & \qquad \qquad \qquad
  P_{Y|XTX'T'K}(\bot|x,t,x',t',k_1) \\
  &  \leq  P_{K|XTX'T'}(k_1|x,t,x',t') \\
  &  = P_{K|XT}(k_1|x,t) = 2^{-|k_1|},
\end{aligned}\end{multline*}
where in the last line we used the fact that knowing one valid
message\-/tag pair $(x,t)$ provides no information about $k_1$,
already expressed in \eqnref{eq:xorhash1}.
\end{proof}

\section{Key recycling for further rounds of authentication}
\label{sec:async}

Since the key recycled in the scheme analyzed in \secref{sec:sync} can
be used for arbitrary applications, it can be used in particular for
authenticating another message. By the composition of the security
definition, running $n$ times that protocol for authenticating $n$
messages is $n\eps$\=/secure and uses a key of total length
$|k_1|+n|k_2|$. This requires synchronization between every message,
e.g., if the timeout is set to $s$ seconds and all messages go from
Alice to Bob with no other communication, then only one message could
be sent every $s$ seconds.

It is however possible to authenticate multiple messages without any
form of synchronization. This is because revealing new message\-/tag
pairs does not leak any information about the key $k_1$ to the
adversary, since the tags are one-time padded by $k_2$. Hence the next
rounds of authentication can be started before Bob receives the
message from the first round. It is only necessary to be synchronized
after all the messages have been delivered, so that $k_1$ may then be
put back in the pool of keys for arbitrary future use.

If the key $k_1$ is deleted after $n$ rounds of authentication instead
of being recycled, then no synchronization is needed at all. The
security proof for the protocol in a completely asynchronous network
follows as a corollary from security with synchronization, in which
$k_1$ is deleted after $n$ rounds.

\subsection{Real and ideal systems}
\label{sec:async.systems}

\begin{figure}[tb]
\begin{centering}

\begin{tikzpicture}[
      resource/.style={draw,minimum width=2.4cm,minimum height=1cm},
      protocol/.style={draw,minimum width=1.3cm,minimum height=2.5cm},
      pnode/.style={minimum width=.8cm,minimum height=.5cm}]

\small

\def\t{3.7} 
\def\u{2.2} 
\def\v{.75}

\node[pnode] (a1) at (-\u,\v) {};
\node[pnode] (a2) at (-\u,0) {};
\node[pnode] (a3) at (-\u,-\v) {};
\node[protocol] (a) at (-\u,0) {};
\node[yshift=-2,above right] at (a.north west) {\footnotesize
  $\pi^{\auth^+_n}_A$};
\node[inner sep=0] (alice1) at (-\t,\v) {};
\node[inner sep=0] (alice2) at (-\t,-\v) {};

\node[pnode] (b1) at (\u,\v) {};
\node[pnode] (b2) at (\u,0) {};
\node[pnode] (b3) at (\u,-\v) {};
\node[protocol] (b) at (\u,0) {};
\node[yshift=-2,above right] at (b.north west) {\footnotesize $\pi^{\auth^+_n}_B$};
\node[inner sep=0] (bob1) at (\t,\v) {};
\node[inner sep=0] (bob2) at (\t,-\v) {};

\node[resource] (keyBox) at (0,\v) {};
\node[draw] (key) at (0,\v) {key};
\node[yshift=-2,above right] at (keyBox.north west) {\footnotesize
  Secret key $\aK_n$};
\node[resource] (channel) at (0,-\v) {};
\node[yshift=-1.5,above] at (channel.north) {\footnotesize
  Insecure channel $\aC_n$};
\node (eveleft) at (-.4,-1.75) {};
\node (everight) at (.4,-1.75) {};
\node (ajunc) at (eveleft |- a3) {};
\node (bjunc) at (everight |- b3) {};

\draw[sArrow] (key) to (a1);
\draw[sArrow] (key) to (b1);

\draw[sArrow] (alice1.center) to  node[auto,pos=.4] {$\{x_i\}_{1}^n$} (a1);
\draw[sArrow] (b1) to node[auto,pos=.6] {$\{y_i\}_{1}^n$} (bob1.center);

\draw[sArrow] (a3) to node[auto,pos=.6,swap] {$k_1$} (alice2.center);
\draw[sArrow] (b3) to node[auto,pos=.6] {$k_1$} (bob2.center);

\draw[sArrow] (a3) to (ajunc.center)
to node[pos=.8,auto,swap] {$\{x_i\|t_i\}_{1}^n$} (eveleft.center);
\draw[sArrow] (everight.center) to node[pos=.2,auto,swap] {$\{x'_i\|t'_i\}_{1}^n$}
(bjunc.center) to (b3);

\end{tikzpicture}

\end{centering}
\caption{\label{fig:multi.real}The \emph{real multiple authentication
    with key recycling} system. Alice has access to the left
  interface, Bob to the right interface and Eve to the lower
  interface. The $\ith{i}$ message $x_i$ is appended with a tag $t_i =
  h_{k_1}(x_i) \xor k^i_2$, where a different $k^i_2$ is used for
  every message but the same $k_1$ is reused throughout. Bob's
  protocol either accepts the received message or outputs an error,
  $y_i \in \{x'_i,\bot\}$.}
\end{figure}
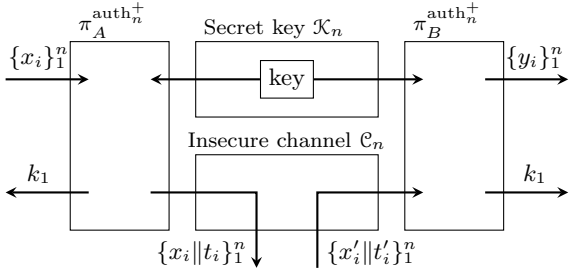

We thus define the protocol $\pi^{\auth_n^+}$ to authenticate $n$
messages by using one key $k_1$, $n$ keys $\{k^i_2\}^n_{i=1}$ \--- we
denote by $\aK_n$ the resource generating these keys \--- and a
multiple use insecure channel $\aC_n$. The tag for the $\ith{i}$
message $x_i$ is given by $t_i = h_{k_1}(x_i) \xor k^i_2$, where
$\{h_{k}\}_{k \in \cK}$ is a $\eps$\=/AXU$_2$ family of hash
functions. As in the previous section, we assume a mild form of
synchronization which allows the key $k_1$ to be recycled by Alice at
the end of the protocol, when all $n$ messages have either been
received or Bob has decided not to accept any more incoming
messages. Bob's protocol can however recycle $k_1$ as soon as all
messages have been received. If some messages are still missing after
the timeout, Bob's protocol outputs an error for each of these
messages. This is depicted in \figref{fig:multi.real}.

\begin{figure}[tb]
\begin{centering}

\begin{tikzpicture}[
      resource/.style={draw,minimum width=3.2cm,minimum height=1.8cm}]

\small

\def\t{2.8}
\def\w{.6}
\def\b{.4}

\def\v{.35}
\def\e{-1.3}

\node[resource] (keyBox) at (0,0) {};
\node[yshift=-1.5,above right] at (keyBox.north west) {\footnotesize
  Authentic channel $\aA^+_n$};
\node[draw] (key) at (0,\v) {key};
\node (a1) at (-\t,\v) {};
\node (a2) at (-\t,-\v) {};
\node (b1) at (\t,\v) {};
\node (b2) at (\t,-\v) {};
\node (e2) at (-\w,\e) {};
\node (e3) at (\w,\e) {};
\node (leak) at (-\w,-\v) {};
\node (s2) at (\w,-\v) {};

\draw[thick] (a2.center) to node[pos=.2,auto] {$\{x_i\}_{i=1}^n$} (\w-\b,-\v) to node[pos=.53]
(handle3) {} +(340:2*\b);
\draw[sArrow] (\w+\b,-\v) to node[pos=.7,auto] {$\{y_i\}_{i=1}^n$} (b2.center);
\draw[sArrow] (key) to node[pos=.75,auto] {$k$} (b1.center);
\draw[sArrow] (key) to node[pos=.75,auto,swap] {$k$} (a1.center);
\draw[sArrow] (leak.center) to node[pos=.8,auto,swap] {$\{x_i\}_{i=1}^n$} (e2.center);
\draw[double] (e3.center) to node[pos=.15,auto,swap] {$\{b_i\}_{i=1}^n
  \in \{0,1\}^n$} (handle3.center);

\end{tikzpicture}

\end{centering}
\caption{\label{fig:multi.resource}A \emph{multiple authentic channel
    with key recycling $\aA^+_n$}. Alice has access to the left
  interface, Bob to the right interface and Eve to the lower
  interface. Alice can send $n$ messages $\{x_i\}_{i=1}^n$. Eve
  obtains these messages, and can choose for each one whether Bob
  receives it or an error $\bot$. After having provided Bob with
  either $y_i \in \{x_i,\bot\}$ for every message, the resource
  generates a new key $k$, which is given to both players.}
\end{figure}
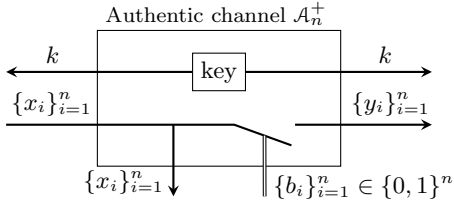

The resource constructed by this protocol is illustrated in
\figref{fig:multi.resource}. It corresponds to a simple modification
of the resource for one message (\figref{fig:recycle.resource}), in
which multiple messages can now be input. For each message, the
adversary can provide one bit $b_i \in \{0,1\}$ that either lets the
original message through or produces an error at Bob's interface. Once
all bits $\{b_i\}_{i=1}^n$ have been input, the resource $\aA^+_n$
outputs a new key $k$ at Bob's interface, and waits for the timeout to
occur to output $k$ at Alice's interface as well.

Note that neither of the resources from Figures~\ref{fig:multi.real}
and \ref{fig:multi.resource} expect all messages $\{x_i\}_{i=1}^n$ to
be input simultaneously, they can be transmitted one at a time. These
messages are processed in the order in which they are received, i.e.,
the first message input at the $A$\=/interface is labeled $x_1$, the
second is labeled $x_2$, etc. Likewise, the $\ith{i}$ message\-/tag
pair input at the $E$\=/interface is labeled $x'_i \| t'_i$ and its
authenticity is verified with the keys $k_1$ and $k^i_2$.


\subsection{Security}
\label{sec:async.security}

To prove that $\pi^{\auth^+_n}$ constructs the ideal resource
$\aA^+_n$ we use as simulator $n$ independent copies of the simulator
$\sigma^{\auth+}_E$ from \figref{fig:recycle.ideal}. We show that with
this simulator, the ideal and real systems are $n\eps$\=/close,
namely \[ \pi^{\auth^+_n} \left(\aK_n \| \aC_n \right) \close{n\eps}
\aA^+_n \sigma^{\auth^+_n}_E.\]

To do this, we rewrite the real and ideal systems as sequences of
boxes which we denote $\aR$ and $\aI$, behaving similarly to one round
of the real and ideal systems, respectively. We then show by induction
that $n$ copies of $\aR$ and one key generation system $\aK$ are
$n\eps$\-/close to $n$ copies of $\aI$ and one $\aK$.

\begin{thm}
  \label{thm:multiple}
  The protocol for authenticating $n$ messages with the same hash
  function, $\pi^{\auth^+_n}$, is $n\eps$\-/secure,
  i.e., $$\aK_n\|\aC_n \xrightarrow{\pi^{\auth^+_n},n\eps} \aA^+_n.$$
\end{thm}

\begin{proof}
  As in the case of one round of authentication from
  \secref{sec:sync}, if Eve is inactive \--- which we model as a
  converters $\sharp_E,\flat_E$ plugged into the $E$\=/interface and
  allowing Alice's messages through \--- the real and ideal
  systems are both equivalent to a channel that perfectly transmits
  $n$ messages. They are thus indistinguishable.

In the case of an active adversary, let $\sigma^{\auth^+_n}_E$ denote
the $n$ copies of the simulator for the one message case, which we use
as simulator in this proof. Consider the real and ideal systems,
namely $\pi^{\auth^+_n} \left(\aK_n \| \aC_n \right)$ and $\aA^+_n
\sigma^{\auth^+_n}_E$. Both of them output at Alice's interface a key
$k_1$ which is a copy of the key output at Bob's interface; both of
them output the key at the same time, when the timeout occurs. Receiving
this key therefore cannot help the distinguisher, so we can ignore
it.

Now consider two boxes. The first, $\aK$, simply produces a uniformly
distributed key $k_1$. The second, $\aR$, behaves similarly to the
composed systems of Alice, Bob and the channels for one round of
authentication. It takes a key $k_1$ and message $x$ as input. If it
receives $k_1$ before $x$, it stores the key, generates a random $k_2$
and waits for $x$. Upon receiving $x$, it outputs $x \| t$ with $t =
h_{k_1}(x) \xor k_2$. If it first receives the message $x$ before
$k_1$, it outputs $x \| t$ where $t$ is a uniformly random tag of the
correct length. Upon receiving $k_1$ later, it retroactively computes
what key $k_2$ would have resulted in the tag $t$, namely $k_2 = t
\xor h_{k_1}(x)$. Upon receiving $x' \| t'$ it checks if $t' =
h_{k_1}(x') \xor k_2$ and outputs either $x'$ or $\bot$ as well as
$k_1$. If no message $x' \| t'$ is input by the timeout, it outputs
$\bot$ and $k_1$.

A sequence of $n$ such boxes \--- depicted on the left in
\figref{fig:recycle.RRII} \--- is identical to the real system without
Alice's recycled key. Both systems output exactly the same messages
with the same probabilities. What is more, every system $\aR$ receives
$k_1$ before $x' \| t'$, because the messages are authenticated in
order, i.e., the $\ith{i}$ pair is sent to the $\ith{i}$ box at which
point $k_1$ is transmitted to the box $i+1$. It can thus always check
whether $t' = h_{k_1}(x') \xor k_2$.

\begin{figure}[tb]
\begin{centering}

\begin{tikzpicture}[
      systemBox/.style={draw,minimum height=1.4cm,minimum width=.7cm},
      invisibleBox/.style={minimum height=1.4cm,minimum width=.7cm},
      keyBox/.style={draw,minimum height=.7cm,minimum width=.7cm}]

\small

\def\b{-1.9}
\def\p{.3}
\def\v{.35} 
\def\vv{.4}
\def\x{4}
\def\y{.7}

\node[keyBox] (Rkey) at (0,\y/2) {$\aK$};
\node[yshift=\b cm,systemBox] (Rsys) at (0,\y) {$\aR$};
\node[yshift=2*\b cm,systemBox] (Rsys2) at (0,\y) {$\aR$};
\node[yshift=3*\b cm,invisibleBox] (Rsys3) at (0,\y) {$\vdots$};
\node[yshift=4*\b cm,systemBox] (Rsys4) at (0,\y) {$\aR$};

\node[systemBox] (Isys) at (\x,0) {$\aI$};
\node[yshift=1*\b cm,systemBox] (Isys2) at (\x,0) {$\aI$};
\node[yshift=2*\b cm,invisibleBox] (Isys3) at (\x,0) {$\vdots$};
\node[yshift=3*\b cm,systemBox] (Isys4) at (\x,0) {$\aI$};
\node[yshift=4*\b cm,keyBox] (Ikey) at (\x,\y/2) {$\aK$};

\draw[sArrow] (Rkey) to node[auto] {$k_1$} (Rsys);
\draw[sArrow] (Rsys) to node[auto] {$k_1$} (Rsys2);
\draw[sArrow] (Rsys2) to node[auto] {$k_1$} (Rsys3);
\draw[sArrow] (Rsys3) to node[auto] {$k_1$} (Rsys4);
\draw[sArrow] (Rsys4) to node[auto] {$k_1$} +(0,-1.2);

\draw[sArrow] (Isys) to node[auto] {\tt ok} (Isys2);
\draw[sArrow] (Isys2) to node[auto] {\tt ok} (Isys3);
\draw[sArrow] (Isys3) to node[auto] {\tt ok} (Isys4);
\draw[sArrow] (Isys4) to node[auto] {\tt ok} (Ikey);
\draw[sArrow] (Ikey) to node[auto] {$k$} +(0,-.85);

\draw[yshift=\b cm,sArrow] (-\v-\vv,\p+\y) to +(\vv,0);
\draw[yshift=\b cm,sArrow] (-\v,-\p+\y) to +(-\vv,0);
\draw[yshift=\b cm,sArrow] (\v,\p+\y) to +(\vv,0);
\draw[yshift=\b cm,sArrow] (\v+\vv,-\p+\y) to +(-\vv,0);

\node[left,yshift=\b cm] at (-\v-\vv,\p+\y) {$x$};
\node[right,yshift=\b cm] at (\v+\vv,\p+\y) {$x\|t$};
\node[right,yshift=\b cm] at (\v+\vv,-\p+\y) {$x'\|t'$};
\node[left,yshift=\b cm] at (-\v-\vv,-\p+\y) {$x',\bot$};

\draw[yshift=\b cm,sArrow] (-\v-\vv+\x,\p) to +(\vv,0);
\draw[yshift=\b cm,sArrow] (-\v+\x,-\p) to +(-\vv,0);
\draw[yshift=\b cm,sArrow] (\v+\x,\p) to +(\vv,0);
\draw[yshift=\b cm,sArrow] (\v+\vv+\x,-\p) to +(-\vv,0);

\node[left,yshift=\b cm] at (-\v-\vv+\x,\p) {$x$};
\node[right,yshift=\b cm] at (\v+\vv+\x,\p) {$x\|t$};
\node[right,yshift=\b cm] at (\v+\vv+\x,-\p) {$x'\|t'$};
\node[left,yshift=\b cm] at (-\v-\vv+\x,-\p) {$x,\bot$};

\draw[yshift=2*\b cm,sArrow] (-\v-\vv,\p+\y) to +(\vv,0);
\draw[yshift=2*\b cm,sArrow] (-\v,-\p+\y) to +(-\vv,0);
\draw[yshift=2*\b cm,sArrow] (\v,\p+\y) to +(\vv,0);
\draw[yshift=2*\b cm,sArrow] (\v+\vv,-\p+\y) to +(-\vv,0);

\node[left,yshift=2*\b cm] at (-\v-\vv,\p+\y) {$x$};
\node[right,yshift=2*\b cm] at (\v+\vv,\p+\y) {$x\|t$};
\node[right,yshift=2*\b cm] at (\v+\vv,-\p+\y) {$x'\|t'$};
\node[left,yshift=2*\b cm] at (-\v-\vv,-\p+\y) {$x',\bot$};

\draw[yshift=3*\b cm,sArrow] (-\v-\vv+\x,\p) to +(\vv,0);
\draw[yshift=3*\b cm,sArrow] (-\v+\x,-\p) to +(-\vv,0);
\draw[yshift=3*\b cm,sArrow] (\v+\x,\p) to +(\vv,0);
\draw[yshift=3*\b cm,sArrow] (\v+\vv+\x,-\p) to +(-\vv,0);

\node[left,yshift=3*\b cm] at (-\v-\vv+\x,\p) {$x$};
\node[right,yshift=3*\b cm] at (\v+\vv+\x,\p) {$x\|t$};
\node[right,yshift=3*\b cm] at (\v+\vv+\x,-\p) {$x'\|t'$};
\node[left,yshift=3*\b cm] at (-\v-\vv+\x,-\p) {$x,\bot$};

\draw[yshift=4*\b cm,sArrow] (-\v-\vv,\p+\y) to +(\vv,0);
\draw[yshift=4*\b cm,sArrow] (-\v,-\p+\y) to +(-\vv,0);
\draw[yshift=4*\b cm,sArrow] (\v,\p+\y) to +(\vv,0);
\draw[yshift=4*\b cm,sArrow] (\v+\vv,-\p+\y) to +(-\vv,0);

\node[left,yshift=4*\b cm] at (-\v-\vv,\p+\y) {$x$};
\node[right,yshift=4*\b cm] at (\v+\vv,\p+\y) {$x\|t$};
\node[right,yshift=4*\b cm] at (\v+\vv,-\p+\y) {$x'\|t'$};
\node[left,yshift=4*\b cm] at (-\v-\vv,-\p+\y) {$x',\bot$};

\draw[sArrow] (-\v-\vv+\x,\p) to +(\vv,0);
\draw[sArrow] (-\v+\x,-\p) to +(-\vv,0);
\draw[sArrow] (\v+\x,\p) to +(\vv,0);
\draw[sArrow] (\v+\vv+\x,-\p) to +(-\vv,0);

\node[left] at (-\v-\vv+\x,\p) {$x$};
\node[right] at (\v+\vv+\x,\p) {$x\|t$};
\node[right] at (\v+\vv+\x,-\p) {$x'\|t'$};
\node[left] at (-\v-\vv+\x,-\p) {$x,\bot$};

\end{tikzpicture}

\end{centering}
\caption{\label{fig:recycle.RRII}The real and ideal authentication
  systems rearranged as a sequence of boxes $\aR$ and $\aI$. $\aI$
  simply generates a uniform tag $t$ independently from the message
  $x$, and outputs an error $\bot$ if $(x\|t) \neq (x'\|t')$ or if no
  message $x$ has been provided. The box $\aR$ transmits $k_1$ to the
  next box only after having output $y \in \{x,\bot\}$. If it receives
  $x$ before having gotten $k_1$ from the previous box, it outputs a
  uniformly random $t$ and retroactively computes $k_2 = h_{k_1}(x)
  \xor t$ when provided with $k_1$.}
\end{figure}
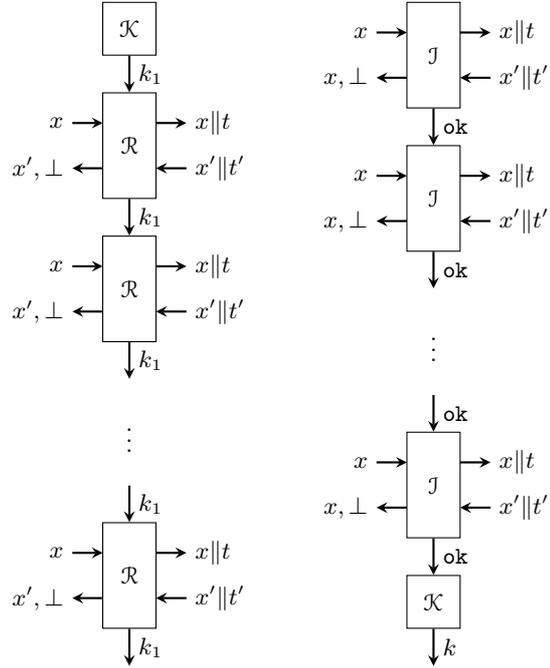

The ideal system can similarly be rewritten as a sequence of boxes
$\aI$, with one final box $\aK$ that generates a new key. After
receiving some message $x$, each box $\aI$ outputs a uniform tag $t$
appended to the message. Upon receiving some $x' \| t'$, each box
outputs $\bot$ if no $x$ had been received or if $(x' \| t') \neq (x
\| t)$, and $x$ otherwise. It then notifies the next box that it is
finished. The final box notifies the key generating box $\aK$, which
outputs a new key. If the timeout occurs, all boxes that have not
produced an output $y \in \{x,\bot\}$ yet, output an error $\bot$ and
the key generation box outputs a new key.

A sequence of $n$ such boxes \--- depicted on the right in
\figref{fig:recycle.RRII} \--- is identical to the ideal system without
Alice's recycled key. Both systems output exactly the same messages
with the same probabilities. So the distance between the real and ideal
systems of multiple message authentication with key recycling is equal
to the distance between these recursive systems depicted in
\figref{fig:recycle.RRII},
\begin{multline*} d\left(\pi^{\auth^+_n} \left(\aK_n \| \aC_n
    \right),\aA^+_n \sigma^{\auth^+_n}_E\right) \\ =
  d\left(\aK\aR_1\dotsb\aR_n,\aI_1\dotsb\aI_n\aK\right)
  ,\end{multline*} where we have numbered the boxes with subscripts.

We prove by induction that these systems can be distinguished with
advantage at most $n\eps$. The case of $n = 1$ is proven by
\thmref{thm:auth}. If this holds for $n-1$, we have
$$d(\aK\aR_1\dotsb\aR_{n-1},\aI_1\dotsb\aI_{n-1}\aK) \leq
(n-1)\eps.$$ Because the distinguishing advantage respects the
triangle inequality (\eqnref{eq:dist.tri}) and is non\-/increasing
under composition with other systems (\eqnref{eq:dist.comp}), we have
  \begin{align*}
    & d(\aK\aR_1\dotsb\aR_n,\aI_1\dotsb\aI_n\aK) \\ & \qquad \leq
    d(\aK\aR_1\dotsb\aR_{n},\aI_1\aK\aR_2\dotsb\aR_{n}) \\ & \qquad
    \qquad \qquad \qquad +
    d(\aI_1\aK\aR_2\dotsb\aR_{n},\aI_1\dotsb\aI_{n}\aK) \\
    & \qquad \leq d(\aK\aR_1,\aI_1\aK) +
    d(\aK\aR_2\dotsb\aR_{n},\aI_2\dotsb\aI_{n}\aK) \\
    & \qquad \leq \eps + (n-1)\eps. \qedhere
  \end{align*}
\end{proof}

The protocol $\pi^{\auth^+_n}$ still uses synchronization for Alice to
know that it is safe to recycle the key $k_1$ for arbitrary use after
the $\ith{n}$ message has been sent. Wegman and Carter's original
authentication scheme does not do this, it only recycles the key for
further rounds of authentication. This can be seen as destroying the
key when there are no more messages to be authenticated. It can thus
be modeled with the real and ideal systems of $\pi^{\auth^+_n}$, but
with the final recycled key removed, which simultaneously removes all
need for synchronization. This can only decrease the distance between
the real and ideal systems. Note that if we remove all form of
synchronization, Bob cannot know that messages have been sent, he will
therefore not output some error $\bot$ if he receives nothing, but
simple wait and do nothing.

\begin{cor}
  \label{cor:wc}
  Wegman and Carter's scheme for authenticating $n$ messages is
  $n\eps$\=/secure in a completely asynchronous network.
\end{cor}

\section{Secret key leakage}
\label{sec:leak}

In this section we look at attacks on Wegman and Carter's
authentication scheme. We show that in the special case where a
$\eps$\=/AXU$_2$ hash function with $\eps = \frac{1}{|\cT|}$ is
recycled for $n$ rounds, there exists an attack which meets the upper
bound on the failure probability of \corref{cor:wc}. This means that
the attack successfully corrupts at least one of the $n$ messages with
probability $n\eps$.

For the adversary to obtain this total success probability, the
success probability in each round must increase, as the following
calculation shows. Let us define $\{F_{i}\}_i$ to be a sequence of
random variables taking the value $1$ if the adversary successfully
corrupts a message in any of the first $i$ rounds, and $0$
otherwise. We then have for any $0 \leq i \leq 1/\eps -
1$, \begin{align*} P_{F_{i+1}|F_{i}}(1|0) & =
  \frac{P_{F_{i+1}}(1)-P_{F_{i+1}F_{i}}(1,1)}{P_{F_{i}}(0)} \\ & =
  \frac{(i+1)\eps-i\eps}{1-i\eps} = \frac{\eps}{1-i\eps}.\end{align*}
$P_{F_{i+1}|F_{i}}(1|0)$ is strictly increasing in $i$, i.e., the
success probability for the adversary increases in every round. This
is because \--- as we show in \thmref{thm:lowerbound} here below \---
some information about the hash function is leaked in every round,
even if the key used for the OTP is perfectly uniform. The entropy of
the hash function gradually decreases, until the adversary has enough
information to successfully corrupt a new message with probability
$1$.

This result contrasts strongly with the non\-/composable analysis found
in~\cite{WC81}. There, the adversary simply collects the pairs of
messages and tags $(x_1,t_1),(x_2,t_2),\dotsc$, and attempts to
corrupt a message in each round, independently from the attempts in
previous rounds. In this case, due to the hiding property of the
OTP, the distribution of the hash function always remains
perfectly uniform given these message\-/tag pairs.

\begin{thm}
\label{thm:lowerbound}
If the same hash function from a family of $\frac{1}{|\cT|}$-almost
XOR universal$_2$ functions is recycled in $n$ rounds of
authentication, for any $1 \leq n \leq |\cT|$, there exists an attack
that allows the adversary to successfully corrupt one of the first $n$
messages with probability at least $n/|\cT|$. Furthermore, after $n$
rounds, the entropy of the recycled key $K$ is bounded
by \[H(K|Z_{n}) \leq \log \frac{|\cK|}{|\cT|} + \left( 1 -
  \frac{n}{|\cT|} \right) \log \left(|\cT| - n \right),\] where
$Z_{n}$ consists of all the inputs and outputs of the protocol \---
the messages, tags, and accept or reject results \--- from these $n$
rounds.
\end{thm}

\begin{proof}
This attack assumes that the adversary, Eve, gets to choose which
messages are sent, and learns whether a corrupted message was
accepted or not. Eve chooses to always send the same message $x$
during these $n$ rounds, and always substitutes the same message $x'
\neq x$ for $x$ in each round as well. To be successful, she needs
to guess correctly the value $c = h_{k_1}(x) \xor h_{k_1}(x')$,
since $t' = t \xor c$, where $t$ is the tag that comes with $x$ and
$t'$ is the correct tag for $x'$. Since the family of hash functions
is $\frac{1}{|\cT|}$-AXU$_2$, the distribution of $c$ is
uniform.\footnote{For any family of functions, $\sum_c \Pr_k \left[
    h_k(x_1) \xor h_k(x_2) = c \right] = 1$ for all $x_1 \neq
  x_2$. So being $\frac{1}{|\cT|}$-AXU$_2$ means that $\Pr_k \left[
    h_k(x_1) \xor h_k(x_2) = c \right] = \frac{1}{|\cT|}$.} Eve
therefore makes a list of the $|\cT|$ possible values for $c$, and
in each round eliminates one from her list.

In the first round she receives $(x,t_1)$ from the insecure channel,
picks a $c_1$ from her list and sends $(x',t_1 \xor c_1)$ back on the
channel. The legitimate player accepts the message received from the
adversary only if $c_1 = h_{k_1}(x) \xor h_{k_1}(x')$, which happens
with probability $p_1 = |\cT|^{-1}$.

If she is unsuccessful at corrupting the message, she can cross $c_1$
off her list. In the second round she then receives $(x,t_2)$, picks a
new $c_2 \neq c_1$, and sends $(x',t_2 \xor c_2)$. This time her
success probability is $p_2 = (|\cT|-1)^{-1}$, since she only has
$|\cT|-1$ elements $c$ left on the list which are all equally
probable.

If we repeat this for each round, the success probability in the
\ith{i} round given that the previous $i-1$ were unsuccessful is $p_i
= (|\cT|-i+1)^{-1}$. We now prove by induction that the probability of
successfully corrupting at least one message with this strategy is
exactly $i/|\cT|$. Let $F_i$ be a random variable taking the value $1$
if the adversary successfully corrupts a message in any of the first
$i$ rounds, and $0$ otherwise. We have $P_{F_1}(1) = p_1 =
1/|\cT|$. And if $P_{F_{i-1}}(1) = (i-1)/|\cT|$, then \begin{align*}
  P_{F_{i}}(1) & =
P_{F_{i-1}}(1) + P_{F_{i-1}}(0)p_i \\ & =
\frac{i-1}{|\cT|}+\left(1-\frac{i-1}{|\cT|}\right)\frac{1}{|\cT|-i+1}
= \frac{i}{|\cT|}.\end{align*}

Let $z_0$ represent any value of $Z_{n}$ in which Eve fails
to corrupt any message, and $z_1$ be the case where she does trick the
legitimate players. If she is successful, she immediately learns the correct value
$c$, and thus \[H(K|Z_n = z_1) = \log \frac{|\cK|}{|\cT|}.\] If
Eve is not successful, she has still managed to cross $n$
values for $c$ off her list, so  
\[H(K|Z_n = z_0) = \log \frac{|\cK|}{|\cT|}\left(|\cT| - n
\right).\]

Combining the two equations above with the corresponding
probabilities, we get
\begin{multline*}H(K|Z_n) = \left( 1 - \frac{n}{|\cT|} \right) \log
  \frac{|\cK|}{|\cT|}\left(|\cT| - n \right) \\ +\frac{n}{|\cT|} \log
  \frac{|\cK|}{|\cT|}. \qedhere\end{multline*}
\end{proof}

\appendix
\appendixpage

\section{Security proof for standard authentication}
\label{app:stdauth}

We showed in \secref{sec:ac} that the security of standard
authentication reduces to proving that \eqnsref{eq:auth.sub} and
\eqref{eq:auth.imp} are bounded by $\eps$ for all choices of input
distributions for $x,x',t'$, which we rewrite here:
\begin{multline}\label{eq:auth.sub.app}
  \frac{1}{2} \sum_{x,t,x',t',y} \big| P_{XTX'T'Y}(x,t,x',t',y) \\ -
    Q_{XTX'T'Y}(x,t,x',t',y) \big| \leq \eps, 
\end{multline}
\begin{multline}\label{eq:auth.imp.app}
  \frac{1}{2} \sum_{x',t',y} \big| P_{X'T'Y}(x',t',y) \\ -
    Q_{X'T'Y}(x',t',y) \big| \leq \eps. 
\end{multline}

In fact, these two equations are strictly weaker than the security
conditions derived by Wegman and Carter~\cite{WC81} and
Stinson~\cite{Sti94}, who maximize over all $x,t,x',t'$ instead of
over $x,x',t'$. It is therefore immediate that previous work on
information\-/theoretic authentication (without key recycling) is
composable. We have however given a simplification of the more common
$\eps$-ASU$_2$ hashing definition, which does not include the extra
requirement that $\Pr \left[h_k(x) = t \right] = \frac{1}{|\cT|}$ (see
\footnoteref{fn:universalhash} on \pref{fn:universalhash}), without
which Stinson's proof~\cite{Sti94} does not apply. We therefore
provide a new proof of security here.

\begin{lem}
  The standard authentication protocol, $\pi^{\auth}$, is $\eps$\=/secure.
\end{lem}

This lemma uses the same proof technique as \thmref{thm:auth}: we reduce
the statistical distance to the sum over all events which are more
probable in the real case, namely accepting a corrupted message.

\begin{proof}
  We start with impersonation attacks, namely
  \eqnref{eq:auth.imp.app}. From the definition of ASU$_2$ hashing, we
  have for any $x',t'$ and $x \neq x'$,
  \begin{align*} \Pr_k \left[ h_k(x') = t' \right] & = \sum_t \left[ h_k(x) = t
    \text{ and } h_k(x') = t' \right] \\ & \leq \eps.\end{align*} And since for all
  $x',t'$ the ideal system always rejects the corrupted message,
  $Q_{Y|X'T'}(\bot|x',t') = 1$, and the maximization of
  \eqnref{eq:auth.imp.app} over $P_{X'T'}$ reduces to
  \begin{align*}
    \max_{x',t'} P_{Y|X'T'}(x'|x',t') = \max_{x',t'} \Pr_k \left[
      h_k(x') = t' \right] \leq \eps.
  \end{align*}

  For the substitution attack we need to show that
  \eqnref{eq:auth.sub.app} is satisfied for all $P_{X} = Q_X$ and
  $P_{X'T'|XT} = Q_{X'T'|XT}$. If the distinguisher chooses $x' = x$,
  both the real and ideal systems behave identically \--- they both
  accept $x$ if $t' = t$ and produce an error otherwise. We can
  therefore only consider the distributions with $P_{XX'}(x,x) =
  0$. In this case, the simulator in the ideal setting always outputs
  an error $\bot$, i.e., $Q_{Y}(\bot) = 1$. Then from \eqnref{eq:sd},
  \eqnref{eq:auth.sub.app} reduces to
\[\sum_{x,t,x',t'} P_{XTX'T'Y}(x,t,x',t',x') \leq \eps.\]

Combining,
\begin{align*}
& P_{XTX'T'}(x,t,x',t') \\ & \qquad \qquad \> = P_{X}(x) P_{T|X}(t|x) P_{X'T'|XT}(x',t'|x,t), \\
& P_{T|X}(t|x) = \Pr_k [ h_k(x) = t], \\
& P_{Y|XTX'T'}(x'|x,t,x',t') \\ & \qquad \qquad \> = \Pr_k [h_k(x') = t' | h_k(x) = t],
\end{align*}
we get
\begin{align*}
& \sum_{x,t,x',t'} P_{XTX'T'Y}(x,t,x',t',x') \\ & \qquad =
\sum_{x,t,x',t'} P_{X}(x) P_{X'T'|XT}(x',t'|x,t) \\ & \qquad \qquad
\qquad \qquad \Pr_k [h_k(x) = t 
 \text{ and } h_k(x') = t'] \\
  & \qquad \leq \sum_{x,t,x',t'} P_{X}(x) P_{X'T'|XT}(x',t'|x,t)
  \frac{\eps}{|\cT|} \\
  & \qquad = \sum_{t} \frac{\eps}{|\cT|} = \eps. \qedhere
\end{align*}
\end{proof}

\section*{Acknowledgments}
\pdfbookmark[1]{Acknowledgments}{ack}

The author would like to thank Renato Renner, Christoph Pacher and
Ueli Maurer for valuable discussions and helpful
comments.

This work has been funded by the Swiss National Science Foundation
(via grant No.~200020-135048 and the National Centre of Competence in
Research `Quantum Science and Technology'), the European Research
Council -- ERC (grant no. 258932) -- and by the Vienna Science and
Technology Fund (WWTF) through project ICT10-067 (HiPANQ).

\newpage

\providecommand{\bibhead}[1]{}
\expandafter\ifx\csname pdfbookmark\endcsname\relax%
  \providecommand{\tocrefpdfbookmark}{}
\else\providecommand{\tocrefpdfbookmark}{%
   \phantomsection%
   \hypertarget{tocreferences}{}%
   \pdfbookmark[1]{References}{tocreferences}}%
\fi

\tocrefpdfbookmark


\begin{thebibliography}{10}

\bibitem{WC81}
Mark~N. Wegman and Larry Carter.
\newblock New hash functions and their use in authentication and set equality.
\newblock {\em Journal of Computer and System Sciences}, 22(3):265--279, 1981.

\bibitem{Sti94}
Douglas~R. Stinson.
\newblock Universal hashing and authentication codes.
\newblock {\em Designs, Codes and Cryptography}, 4(3):369--380, 1994.
\newblock A preliminary version appeared at CRYPTO~'91.
\newblock [\epfmtdoi{10.1007/BF01388651}].

\bibitem{Can01}
Ran Canetti.
\newblock Universally composable security: A new paradigm for cryptographic
  protocols.
\newblock In {\em Proceedings of the 42nd Symposium on Foundations of Computer
  Science, FOCS~'01}, pages 136--145. IEEE, 2001.
\newblock [\epfmtdoi{0.1109/SFCS.2001.959888}].

\bibitem{Can13}
Ran Canetti.
\newblock Universally composable security: A new paradigm for cryptographic
  protocols.
\newblock Cryptology ePrint Archive, Report 2000/067, 2013.
\newblock Updated version of~\cite{Can01}.
\newblock [\epfmt{cryptoeprint}{2000/067}].

\bibitem{CDPW07}
Ran Canetti, Yevgeniy Dodis, Rafael Pass, and Shabsi Walfish.
\newblock Universally composable security with global setup.
\newblock In {\em Theory of Cryptography, Proceedings of TCC 2007}, volume 4392
  of {\em Lecture Notes in Computer Science}, pages 61--85. Springer, 2007.
\newblock [\epfmtdoi{10.1007/978-3-540-70936-7_4},
  \epfmt{cryptoeprint}{2006/432}].

\bibitem{PW00}
Birgit Pfitzmann and Michael Waidner.
\newblock Composition and integrity preservation of secure reactive systems.
\newblock In {\em Proceedings of the 7th ACM Conference on Computer and
  Communications Security, CSS~'00}, pages 245--254. ACM, 2000.
\newblock [\epfmtdoi{10.1145/352600.352639}].

\bibitem{PW01}
Birgit Pfitzmann and Michael Waidner.
\newblock A model for asynchronous reactive systems and its application to
  secure message transmission.
\newblock In {\em IEEE Symposium on Security and Privacy}, pages 184--200.
  IEEE, 2001.
\newblock [\epfmtdoi{10.1109/SECPRI.2001.924298}].

\bibitem{BPW04}
Michael Backes, Birgit Pfitzmann, and Michael Waidner.
\newblock A general composition theorem for secure reactive systems.
\newblock In {\em Theory of Cryptography, Proceedings of TCC 2004}, volume 2951
  of {\em Lecture Notes in Computer Science}, pages 336--354. Springer, 2004.
\newblock [\epfmtdoi{10.1007/978-3-540-24638-1_19}].

\bibitem{BPW07}
Michael Backes, Birgit Pfitzmann, and Michael Waidner.
\newblock The reactive simulatability ({RSIM}) framework for asynchronous
  systems.
\newblock {\em Information and Computation}, 205(12):1685--1720, 2007.
\newblock Extended version of~\cite{PW01}.
\newblock [\epfmtdoi{10.1016/j.ic.2007.05.002},
  \epfmt{cryptoeprint}{2004/082}].

\bibitem{BM04}
Michael {Ben-Or} and Dominic Mayers.
\newblock General security definition and composability for quantum \&
  classical protocols.
\newblock eprint, 2004.
\newblock [\epfmt{arxiv}{quant-ph/0409062}].

\bibitem{Unr04}
Dominique Unruh.
\newblock Simulatable security for quantum protocols.
\newblock eprint, 2004.
\newblock [\epfmt{arxiv}{quant-ph/0409125}].

\bibitem{Unr10}
Dominique Unruh.
\newblock Universally composable quantum multi-party computation.
\newblock In {\em Advances in Cryptology -- EUROCRYPT 2010}, volume 6110 of
  {\em Lecture Notes in Computer Science}, pages 486--505. Springer, 2010.
\newblock [\epfmtdoi{10.1007/978-3-642-13190-5_25}, \epfmt{arxiv}{0910.2912}].

\bibitem{MR11}
Ueli Maurer and Renato Renner.
\newblock Abstract cryptography.
\newblock In {\em Proceedings of Innovations in Computer Science, ICS 2010},
  pages 1--21. Tsinghua University Press, 2011.

\bibitem{SBCDLP09}
Valerio Scarani, Helle Bechmann-Pasquinucci, Nicolas~J. Cerf, Miloslav
  Du\ifmmode~\check{s}\else \v{s}\fi{}ek, Norbert L\"utkenhaus, and Momtchil
  Peev.
\newblock The security of practical quantum key distribution.
\newblock {\em Reviews of Modern Physics}, 81:1301--1350, September 2009.
\newblock [\epfmtdoi{10.1103/RevModPhys.81.1301}, \epfmt{arxiv}{0802.4155}].

\bibitem{BHLMO05}
Michael {Ben-Or}, Michael Horodecki, Debbie Leung, Dominic Mayers, and Jonathan
  Oppenheim.
\newblock The universal composable security of quantum key distribution.
\newblock In {\em Theory of Cryptography, Proceedings of TCC 2005}, volume 3378
  of {\em Lecture Notes in Computer Science}, pages 386--406. Springer, 2005.
\newblock [\epfmtdoi{10.1007/978-3-540-30576-7_21},
  \epfmt{arxiv}{quant-ph/0409078}].

\bibitem{MR09}
J\"orn M\"uller-Quade and Renato Renner.
\newblock Composability in quantum cryptography.
\newblock {\em New Journal of Physics}, 11(8):085006, 2009.
\newblock [\epfmtdoi{10.1088/1367-2630/11/8/085006}, \epfmt{arxiv}{1006.2215}].

\bibitem{PR14}
Christopher Portmann and Renato Renner.
\newblock Cryptographic security of quantum key distribution.
\newblock eprint, 2014.
\newblock [\epfmt{arxiv}{1409.3525}].

\bibitem{Kra94}
Hugo Krawczyk.
\newblock {LFSR}-based hashing and authentication.
\newblock In {\em Advances in Cryptology -- CRYPTO~'94}, volume 839 of {\em
  Lecture Notes in Computer Science}, pages 129--139. Springer, 1994.
\newblock [\epfmtdoi{10.1007/3-540-48658-5_15}].

\bibitem{Kra95}
Hugo Krawczyk.
\newblock New hash functions for message authentication.
\newblock In {\em Advances in Cryptology -- EUROCRYPT~'95}, volume 921 of {\em
  Lecture Notes in Computer Science}, pages 301--310. Springer, 1995.
\newblock [\epfmtdoi{10.1007/3-540-49264-X_24}].

\bibitem{Rog99}
Phillip Rogaway.
\newblock Bucket hashing and its application to fast message authentication.
\newblock {\em Journal of Cryptology}, 12(2):91--115, 1999.
\newblock A preliminary version appeared at CRYPTO~'95.
\newblock [\epfmtdoi{10.1007/PL00003822}].

\bibitem{AS96}
Mustafa Atici and Douglas~R. Stinson.
\newblock Universal hashing and multiple authentication.
\newblock In {\em Advances in Cryptology -- CRYPTO~'96}, volume 1109 of {\em
  Lecture Notes in Computer Science}, pages 16--30. Springer, 1996.
\newblock [\epfmtdoi{10.1007/3-540-68697-5_2}].

\bibitem{HLM11}
Patrick Hayden, Debbie Leung, and Dominic Mayers.
\newblock The universal composable security of quantum message authentication
  with key recycling.
\newblock Presented at QCrypt 2011, 2011.

\bibitem{Sho96}
Victor Shoup.
\newblock On fast and provably secure message authentication based on universal
  hashing.
\newblock In {\em Advances in Cryptology -- CRYPTO~'96}, volume 1109 of {\em
  Lecture Notes in Computer Science}, pages 313--328. Springer, 1996.
\newblock [\epfmtdoi{10.1007/3-540-68697-5_24}].

\bibitem{Ber05}
Daniel~J. Bernstein.
\newblock Stronger security bounds for wegman-carter-shoup authenticators.
\newblock In {\em Advances in Cryptology -- EUROCRYPT 2005}, volume 3494 of
  {\em Lecture Notes in Computer Science}, pages 164--180. Springer, 2005.
\newblock [\epfmtdoi{10.1007/11426639_10}].

\bibitem{Can04}
Ran Canetti.
\newblock Universally composable signature, certification, and authentication.
\newblock In {\em Proceedings of the 17th IEEE Computer Security Foundations
  Workshop}, page 219. IEEE, 2004.
\newblock [\epfmtdoi{10.1109/CSFW.2004.24}, \epfmt{cryptoeprint}{2003/239}].

\bibitem{CL08}
J\"{o}rgen Cederl\"{o}f and Jan{-}\r{A}ke Larsson.
\newblock Security aspects of the authentication used in quantum cryptography.
\newblock {\em IEEE Transactions on Information Theory}, 54(4):1735--1741,
  2008.
\newblock [\epfmtdoi{10.1109/TIT.2008.917697},
  \epfmt{arxiv}{quant-ph/0611009}].

\bibitem{AL11}
Aysajan Abidin and Jan{-}\r{A}ke Larsson.
\newblock Security of authentication with a fixed key in quantum key
  distribution.
\newblock eprint, 2011.
\newblock [\epfmt{arxiv}{1109.5168}].

\bibitem{Mau13}
Ueli Maurer.
\newblock Authentication amplification by synchronization.
\newblock In {\em Proceedings of the 2013 IEEE International Symposium on
  Information Theory, ISIT 2013}, pages 2711--2714. IEEE, 2013.
\newblock [\epfmtdoi{10.1109/ISIT.2013.6620719}].

\bibitem{Wal14}
Nino Walenta, Andreas Burg, Dario Caselunghe, Jeremy Constantin, Nicolas Gisin,
  Olivier Guinnard, Raphael Houlmann, Pascal Junod, Boris Korzh, Natalia
  Kulesza, Matthieu Legr\'e, Charles Ci~Wen Lim, Tommaso Lunghi, Laurent Monat,
  Christopher Portmann, Mathilde Soucarros, Patrick Trinkler, Gregory Trolliet,
  Fabien Vannel, and Hugo Zbinden.
\newblock A fast and versatile quantum key distribution system with hardware
  key distillation and wavelength multiplexing.
\newblock {\em New Journal of Physics}, 16(1):013047, 2014.
\newblock [\epfmtdoi{10.1088/1367-2630/16/1/013047}, \epfmt{arxiv}{1309.2583}].

\bibitem{BJKS94}
J\"{u}rgen Bierbrauer, Thomas Johansson, Gregory Kabatianskii, and Ben Smeets.
\newblock On families of hash functions via geometric codes and concatenation.
\newblock In {\em Advances in Cryptology -- CRYPTO~'93}, volume 773 of {\em
  Lecture Notes in Computer Science}, pages 331--342. Springer, 1994.
\newblock [\epfmtdoi{10.1007/3-540-48329-2_28}].

\bibitem{Sti95}
Douglas~R. Stinson.
\newblock On the connections between universal hashing, combinatorial designs
  and error-correcting codes.
\newblock {\em Electronic Colloquium on Computational Complexity (ECCC)},
  2(52), 1995.

\bibitem{Mau12}
Ueli Maurer.
\newblock Constructive cryptography---a new paradigm for security definitions
  and proofs.
\newblock In {\em Proceedings of Theory of Security and Applications, TOSCA
  2011}, volume 6993 of {\em Lecture Notes in Computer Science}, pages 33--56.
  Springer, 2012.
\newblock [\epfmtdoi{10.1007/978-3-642-27375-9_3}].

\bibitem{DFPR14}
Vedran Dunjko, Joseph Fitzsimons, Christopher Portmann, and Renato Renner.
\newblock Composable security of delegated quantum computation.
\newblock To appear at ASIACRYPT 2014, 2014.
\newblock [\epfmt{arxiv}{1301.3662}].

\bibitem{Mau02}
Ueli Maurer.
\newblock Indistinguishability of random systems.
\newblock In Lars Knudsen, editor, {\em Advances in Cryptology -- EUROCRYPT
  2002}, volume 2332 of {\em Lecture Notes in Computer Science}, pages
  110--132. Springer, 2002.
\newblock [\epfmtdoi{10.1007/3-540-46035-7_8}].

\bibitem{MPR07}
Ueli Maurer, Krzysztof Pietrzak, and Renato Renner.
\newblock Indistinguishability amplification.
\newblock In {\em Advances in Cryptology -- CRYPTO 2007}, volume 4622 of {\em
  Lecture Notes in Computer Science}, pages 130--149. Springer, 2007.
\newblock [\epfmtdoi{10.1007/978-3-540-74143-5_8}].

\bibitem{GW07}
Gus Gutoski and John Watrous.
\newblock Toward a general theory of quantum games.
\newblock In {\em Proceedings of the 39th Symposium on Theory of Computing,
  STOC~'07}, pages 565--574. ACM, 2007.
\newblock [\epfmtdoi{10.1145/1250790.1250873}].

\bibitem{CDP09}
Giulio Chiribella, Giacomo~Mauro D'Ariano, and Paolo Perinotti.
\newblock Theoretical framework for quantum networks.
\newblock {\em Physical Review A}, 80:022339, August 2009.
\newblock [\epfmtdoi{10.1103/PhysRevA.80.022339}, \epfmt{arxiv}{0904.4483}].

\end{thebibliography}
\end{document}